\documentclass[a4paper,12pt]{article}
\pdfoutput=1
\usepackage[pdftex,unicode,
            colorlinks=true,
          linkcolor = blue]{hyperref}
\usepackage[T1, T2A]{fontenc}
\usepackage[utf8]{inputenc}
\usepackage[english]{babel}
\usepackage{amssymb, amsmath}

\usepackage{geometry}

\usepackage[onehalfspacing]{setspace}

\numberwithin{equation}{section}

\title{}
\author{}
\date{}

\begin{document}

\begin{center}
{\Large \bf Superfield realization of hidden $R$-symmetry \\
\vspace{0.2cm} in extended supersymmetric gauge theories and its applications}
\end{center}
\vspace{1cm}

\begin{center}

{\bf I.L. Buchbinder${}^{a}$\footnote{joseph@tspu.edu.ru}, E.A.
Ivanov${}^b$\footnote{eivanov@theor.jinr.ru}, V.A.
Ivanovskiy${}^c$\footnote{ivanovskiy.va@phystech.edu}} \vspace{5mm}

\footnotesize{ ${}^{a}${\it Department of Theoretical Physics,
Tomsk State Pedagogical University, 634061, Tomsk, Russia} \\
and {\it National Research Tomsk State University, Tomsk, Russia}\\
 ${}^{b}$ {\it Bogoliubov Laboratory of Theoretical
Physics, JINR,
141980 Dubna, Moscow region, Russia}\\
${}^{c}$ {\it Moscow Institute of Physics and Technology, 1417000 Dolgoprudny, Russia}
}

\vspace{1.5cm}

\end{center}

\begin{abstract}
We present the explicit superfield realizations of the hidden
$SU(4)$ and $O(5)$  $R$-symmetries in $4D,\, {\cal N}=4$ and $5D,\,
{\cal N}=2$ supersymmetric Yang-Mills theories in the harmonic
superspace approach. The $R$-symmetry transformations are
constructed and their algebraic structure is studied.  It is shown
that such transformations are consistent with both manifest and
hidden supersymmetry transformations. These symmetries can serve as
an alternative tool for constructing the relevant complete
low-energy superfield effective actions determined earlier from the
hidden supersymmetry considerations.
\end{abstract}
\vspace{1.5cm}

\thispagestyle{empty}
\vfill

\setcounter{page}{0} \setcounter{footnote}{0}

\section{Introduction}
Classical and quantum supersymmetric Yang-Mills (SYM) theories with 16
supercharges in diverse dimensions play an important role in the modern
field theory(see, e.g., \cite{Seiberg}). The main source of interest in them is the property
that they describe low-energy limit of some
brane-like compactifications of type II string theory and thereby
provide a bridge between superstring theory and supersymmetric field
theory. Most interesting and elaborated are $4D, {\cal N}=4$, $5D,\,
{\cal N}=2$ and $6D, \,{\cal N} =(1,1)$ SYM theories. Some of these
models admit superfield formulations with half of the underlying
supersymmetry being manifest and off-shell, namely, with $4D,\, {\cal
N}=2,\,$  $5D, \,{\cal N}=1$ and $6D,\, {\cal N} =(1,0)$ off-shell
supersymmetries. These formulations were constructed within the
relevant harmonic superspaces \cite{litlink 10,litlink 1}.

In such formulations, the second half of the total supersymmetry is
realized on the basic superfields of the theory as a hidden (or
implicit) supersymmetry which forms a closed Lie bracket structure
with the manifest supersymmetry only on shell. While inspecting the
superfield quantum effective actions, the role of this hidden half
of supersymmetry turns out to be very restrictive: requiring the
effective action to enjoy such a supersymmetry fixes, in most known
cases,  its structure up to an overall coefficient which should further
be explicitly calculated from the relevant superfield quantum
perturbation theory. A review of this approach is given in Refs.
\cite{litlink 23, litlink 22, BI}.

The basic example of applying such a strategy for constructing the
quantum effective action is provided by $4D, \,\mathcal{ N}=4$ SYM
effective action which was constructed in \cite{litlink 8} as a hypermultiplet
completion of the non-holomorphic ${\cal N}=2$ gauge superfield
potential found in \cite{litlink 2}. Later on, the same effective action
was reproduced in  various harmonic superspaces \cite{litlink 22}. These works revealed a correspondence between the
$\mathcal{ N} = 4$ SYM low-energy effective action and the leading
terms in the effective action of D3 brane on the $AdS^5 \times S^5$
background. One more example is $5D,\, \mathcal{ N}=2$
SYM theory. Its low-energy effective action was constructed as a
hypermultiplet completion  of the $5D, \,\mathcal{ N}=1$ SYM effective action
\cite{litlink 5}.

In this parer we propose another way to determine the low-energy
effective actions of $4D$, $\mathcal{ N}=4$ and $5D$, $\mathcal{
N}=2$ SYM theories in harmonic superspace. It is based upon exploiting a hidden
bosonic $R$-symmetry of these theories instead of hidden supersymmetry.
To be more precise, we suggest a new
possibility to construct the superspace functionals depending on all
fields of the corresponding supermultiplet, beginning with a functional
which involves only part of such fields. The point is
that the supersymmetry algebra possesses the automorphism group
which is called the $R$-symmetry group. We will show that such
$R$-symmetry can be realized directly on harmonic superfields. While some subgroups of $R$-symmetry
are realized linearly and manifestly, the rest of its transformations proves to possess a highly non-trivial
realization mixing all fields of the extended supermultiplet.  As a result, we gain a possibility to impose the condition of invariance under
the total $R$-symmetry on a superspace functional in order to specify its dependence
on all fields of such a  supermultiplet.

In the harmonic superspace formulation, the full multiplets of $4D$,
$\mathcal{ N}=4$ or  $5D$, $\mathcal{ N}=2$ SYM theories consist of
the gauge vector multiplet and the hypermultiplet. Not only half of $4D, \,\mathcal{ N}=4$ and $5D,\, \mathcal{ N}=2$ supersymmetries is
realized in an implicit way, but also that part of the total
$R$-symmetries of these theories, viz., of $SU(4)$ and $SO(5)$, which mixes the hidden and manifest supersymmetry
transformations. It is also realized by some implicit transformations.
In this paper we find the precise form of the hidden $R$-symmetry
transformations which extend the manifest $R$-symmetry groups,
namely $U(2) \times SU(2)$ and $SU(2) \times SU(2)$,  to $SO(6)$ or
$SO(5)$, respectively.

Although  the low-energy effective action might be found by direct
quantum computations in harmonic superspace or by using the hidden
supersymmetry transformations, we determine it here in a different way.
Namely, we construct the hypermultiplet completions of  $4D, \,\mathcal{ N}=2$ and
$5D,\, \mathcal{ N}=1$ leading terms  by imposing the requirement of full
$R$-symmetry invariance. The effective action corresponds to the
Coulomb branch, with the gauge group being broken to some abelian
subgroup. For simplicity we concentrate on $SU(2)$ gauge group
broken to $U(1)$ and consider, as usual in the Coulomb phase, only that
part of the effective action which depends on the fields of
massless $U(1)$ ${\cal N}=2$ gauge multiplet and its  neutral
hypermultiplet partner forming together $4D, \,\mathcal{ N}=4$ or $5D,\,
\mathcal{ N}=2$ abelian $U(1)$ gauge multiplets.

In addition, we explicitly show that the $4D, \, \mathcal{ N}=4$ SYM
effective action respects superconformal invariance.

\section{$4D, \mathcal{ N}=4$  SYM theory}

In this section we present the superfield realization of the hidden part of the $R$-symmetry
transformations for $4D, \, \mathcal{N} = 4$ SYM theory with the gauge
group $SU(2)$ and construct the low-energy effective action. As an additional exercise,
we show its superconformal invariance.

\subsection{A sketch of ${\cal N}=2$ harmonic superspace}
In our presentation we basically follow the notation and conventions
of Refs. \cite{litlink 1, litlink 22}. Some important notions of the harmonic superspace approach
are briefly outlined below.

The standard $4D, \,\mathcal{ N}=2$ superspace is parametrized by the coordinates
\begin{equation}
z^M=(x^m, \theta^{\alpha}_i, \bar\theta^{\dot{\alpha}i} ),
\end{equation}
where $x^m$, $m=0,1,2,3\,$,   are the Minkowski space coordinates and $\theta^{\alpha}_i$, $\bar\theta^{\dot{\alpha}i}$,
$i=1,2 $, $\alpha,\dot\alpha=1,2$, are the anticommuting Grassmann coordinates.

One can add, to this set of coordinates, the harmonics  $u^{\pm i}$ ($u^{-}_i=(u^{+i})^*$, $u^{+i}u^{-}_i=1$) which describe the ``harmonic sphere''
$SU (2)_R/U (1)$, where $SU (2)_R$ is the $R$-symmetry group acting on the doublet indices $i, k\,$.
The  $4D,\, \mathcal{ N} = 2$  harmonic  superspace in the central basis is
defined as the enlarged coordinate set
\begin{equation}
Z=(z, u)=(x^m, \theta^{\alpha}_i, \bar\theta^{\dot{\alpha}i}, u^{\pm i}).
\end{equation}
In the analytic basis it is parametrized by the coordinates
\begin{gather}
Z_{(an)}=(x^m_{(an)}, \theta^{\pm}_{\alpha}, \bar\theta^{\pm}_{\dot{\alpha}}, u^{\pm i}), \label{HSSa} \\
x^m_{an}=x^m-2i\theta^{(i}\sigma^m\bar\theta^{j)}u^+_iu^-_j, \quad\theta^{\pm}_\alpha=u^{\pm}_i\theta^{i}_\alpha, \qquad\bar\theta^{\pm}_{\dot\alpha}=u^{\pm}_i\bar\theta^{i}_{\dot\alpha}.
\end{gather}
The most important feature of the analytic basis consists in that the set of coordinates
\begin{equation}
\zeta=(x^m_{(an)}, \theta^{+}_{\alpha}, \bar\theta^{+}_{\dot{\alpha}}, u^{\pm i}),
\label{11111}
\end{equation}
involving only half of the original Grassmann coordinates, forms a subspace closed under the  $4D$, $\mathcal{ N}=2$, supersymmetry transformations.
The set (\ref{11111}) represents what is called the ``harmonic analytic superspace''.

The important ingredients of the harmonic superspace approach are the spinor and harmonic derivatives.
In the analytic basis, they are expressed as
\begin{equation}
\begin{split}
&D^{+}_\alpha=\frac{\partial}{\partial \theta^{-\alpha}}, \qquad
\bar D^{+}_{\dot\alpha}=\frac{\partial}{\partial\bar \theta^{-\dot\alpha}},\\
&D^{-}_\alpha=-\frac{\partial}{\partial \theta^{+\alpha}}+2i\bar \theta^{-\dot\alpha}\partial_{\alpha\dot \alpha}, \qquad
\bar D^{-}_{\dot\alpha}=-\frac{\partial}{\partial \theta^{+\dot\alpha}}+2i \theta^{-\alpha}\partial_{\alpha\dot \alpha},\\
&D^{++}=u^{+i}\frac{\partial}{\partial u^{-i}}-2i\theta^{+\alpha}\bar\theta^{+\dot\alpha}\partial_{\alpha\dot \alpha}
+\theta^{+\alpha}\frac{\partial}{\partial \theta^{-\alpha}}+\bar\theta^{+\dot\alpha}\frac{\partial}{\partial \bar\theta^{-\dot\alpha}},\\
&D^{--}=u^{-i}\frac{\partial}{\partial u^{+i}} -2i\theta^{-\alpha}\bar\theta^{-\dot\alpha}\partial_{\alpha\dot \alpha}+\theta^{-\alpha}
\frac{\partial}{\partial \theta^{+\alpha}}+\bar\theta^{-\dot\alpha}\frac{\partial}{\partial \bar\theta^{+\dot\alpha}}.
\label{DeriVat}
\end{split}
\end{equation}
The harmonic derivatives $D^{\pm\pm}$, together with the harmonic $U(1)$ charge operator
$$
D^0 = u^{+i}\frac{\partial}{\partial u^{+i}} - u^{-i}\frac{\partial}{\partial u^{-i}} +
\theta^{+\alpha}\frac{\partial}{\partial \theta^{+\alpha}}+\bar\theta^{+\dot\alpha}\frac{\partial}{\partial \bar\theta^{+\dot\alpha}} -
\theta^{-\alpha}
\frac{\partial}{\partial \theta^{-\alpha}} -\bar\theta^{-\dot\alpha}\frac{\partial}{\partial \bar\theta^{-\dot\alpha}}\,,
$$
form an $SU(2)$ algebra,
\begin{equation}
[D^{++}, D^{--}] = D^0\,, \quad [ D^0, D^{\pm\pm}] = \pm 2 D^{\pm\pm}\,.
\end{equation}
The harmonic superfields (as well as the harmonic projections of the spinor covariant derivatives) carry a definite
integer harmonic $U(1)$ charge, $D^0 \Phi^{q}(Z) = q \Phi^{q}(Z), \,[D^0, D^\pm_{\alpha, \dot\alpha}] = \pm D^\pm_{\alpha, \dot\alpha} $.
This harmonic charge  is assumed to be strictly preserved in any superfield action defined on the superspaces (\ref{HSSa}) or (\ref{11111}).

The ``shortness'' of the spinor derivatives $D^{+}_\alpha,\bar D^{+}_{\dot\alpha}$ in (\ref{DeriVat}) reflects the existence
of the analytic harmonic subspace (\ref{11111}) in the general harmonic superspace (\ref{HSSa}): one can define an analytic ${\cal N}=2$
superfield by imposing the proper covariant ``Grassmann analyticity'' constraints on a general harmonic superfield, viz., $D^{+}_\alpha \Phi^{q}(Z) =
\bar D^{+}_{\dot\alpha}\Phi^{q}(Z) = 0 \;\Rightarrow \;\Phi^{q}(Z) = \phi^{q}(\zeta)\,.$ The harmonic derivative $D^{++}$ commutes with these spinor derivatives and
so preserves the Grassmann harmonic analyticity: $D^{++}\Phi^{q}(Z)$ is an analytic superfield if $ \Phi^{q}(Z)$ is.

\subsection{Classical action of ${\cal N}=4$ SYM}

When formulated in $\mathcal{ N}=2$ harmonic superspace, $\mathcal{ N}=4$
vector gauge  multiplet can be viewed as a ``direct sum'' of the gauge $\mathcal{ N}=2$
multiplet and the hypermultiplet described, respectively, by the analytic superfields
$V^{++}(\zeta)$ and $q^{+}_a(\zeta) =(q^+(\zeta), -\tilde
q^{+}(\zeta)),$ where ``tilde'' means some generalized complex conjugation \cite{litlink 1}.
Both these multiplets belong to the same adjoint
representation of the gauge group. The $\mathcal{ N}=2$ gauge
multiplet  $V^{++}$ is described dy the classical action \cite{litlink 9}
\begin{equation}
S^{
\mathcal{N}=2}_{\text{SYM}}=\frac{1}{2}\sum\limits_{n=2}^{\infty}\text{tr}\frac{(-i)^n}{n}\int
d^{12}z du_1\dots du_n\frac{V^{++}(z,u_1)\dots
V^{++}(z,u_n)}{(u^+_1u^+_2)\dots(u^+_nu^+_1)},
\label{0001}
\end{equation}
where integration goes over the total harmonic superspace and the harmonic distributions $1/(u^+_1u^+_2), \cdots$ are defined in
\cite{litlink 1}.

This action yields the following equations of motion
\begin{equation}
(D^{+})^2W=0, \qquad (\bar{D}^{+})^2\bar W=0,
\end{equation}
where $(D^{+})^2= D^{+\alpha}D^{+}_\alpha$, $(\bar{D}^{+})^2= \bar{D}^{+}_{\dot\alpha}\bar{D}^{+\dot\alpha}$ and $D^+_\alpha, \bar{D}^+_{\dot\alpha}$ are the harmonic plus-projection
of the gauge-covariant spinor derivatives in the so called ``$\lambda$'' frame, in which these derivatives require no gauge connection terms and
coincide with their flat counterparts defined in (\ref{DeriVat}), $W$ and $\bar W$ are chiral and antichiral gauge superfield strengths. The latter can de expressed in terms of
the non-analytic harmonic gauge connection $V^{--}\,,$
\begin{equation}
W=-\frac{1}{4}(\bar D^{+})^2V^{--}, \qquad  \bar
W=-\frac{1}{4}(D^{+})^2V^{--}, \label{DefWW}
\end{equation}
where $V^{--}$ is related to $V^{++}$ by the harmonic flatness
condition
\begin{equation}
D^{--}V^{++}-D^{++}V^{--}+i[V^{++},V^{--}]=0. \label{0004}
\end{equation}

    The classical action for the analytic hypermultiplet in the adjoint representation  reads \cite{litlink 10}
    \begin{equation}
    S_q=\frac{1}{2}\text{tr}\int d\zeta^{-4}q^{+}_a\nabla^{++}q^{+a}=
    \frac{1}{2}\text{tr}\int d\zeta^{-4}q^{+}_a\left(D^{++}q^{+a}+i[V^{++},q^{+a}]\right),
    \label{0005}
    \end{equation}
    where $ d\zeta^{-4}$ is the measure of integration over the analytic harmonic superspace. This action is invariant
    under an extra $SU(2)_{\text{PG}}$ symmetry  transforming $q^{+a}$ as a doublet. Both actions (\ref{0001}) and (\ref{0005}) are
    invariant under the standard linear automorphism group $SU(2)_R$ which rotates the doublet indices of the harmonic variables. In addition, both
    actions are invariant under the separate $R$-symmetry $U(1)_R$ which transforms $\theta^\pm$ and $\bar\theta^\pm$ by the conjugated phase factors.
    Correspondingly, $W$ and $\bar{W}$ defined in (\ref{DefWW}) are also transformed by the appropriate mutually conjugated phase factors, $q^{+ a}$ is the $U(1)_R$ singlet.

    The action of $\mathcal{ N}=4$ SYM theory in ${\cal N}=2$ harmonic superspace is the sum of the actions  (\ref{0001}) and (\ref{0005}),
    \begin{equation}
    S^{ \mathcal{N}=4}_{\text{SYM}}=S^{ \mathcal{N}=2}_{\text{SYM}}+S_q.
    \label{2.40}
    \end{equation}
    The total action is invariant under the following hidden  $\mathcal{ N}=2$ supersymmetry transformations which
    complement the manifest $\mathcal {N} = 2 $ supersymmetry to the full $\mathcal {N} = 4 $ supersymmetry
    \begin{eqnarray}
    \delta V^{++}=\left[\epsilon^{a\alpha}\theta^{ +}_{\alpha}-\bar\epsilon^a_{\dot\alpha}\bar\theta^{ +\dot\alpha}\right]q^{+}_{a}, \qquad
    \delta q^{+}_a=-\frac{1}{32}(D^{+})^2(\bar D^{+})^2\left[\epsilon^{\alpha}_a\theta^{-}_\alpha V^{--}+\bar\epsilon^{}_{a\dot \alpha}\bar\theta^{-\dot\alpha} V^{--}\right],
    \label{00.5}
    \end{eqnarray}
    with $\bar\epsilon^{}_{a\dot \alpha}$ and $\epsilon^{\alpha}_a$ as new anticommuting parameters.
    Though checking the invariance of the action does not require the use of the classical equations of motion, the algebra of these transformations is closed modulo terms proportional to the equations of motion.
    Therefore, in this formulation only the manifest $\mathcal{ N}=2$ supersymmetry is off-shell closed.

\subsection{$R$-symmetry transformations}
We define the additional  $R$-symmetry transformations of the gauge and hypermultiplet harmonic
superfields as follows
\begin{equation}
\begin{split}
&\delta V^{++}=\left[\lambda^{-a}(\theta^{ +})^2+\bar\lambda^{-a}(\bar\theta^{ +})^2\right]q^{+}_{a},\\
&\delta q^{+}_a=\frac{(D^{+})^2(\bar
    D^{+})^2}{64}\left[\lambda^{+}_a(\theta^{-})^2
V^{--}-2\lambda^{-}_a\theta^{+\alpha}\theta^{-}_{\alpha}
V^{--}+\bar\lambda^{+}_{a}(\bar\theta^{-})^2
V^{--}-2\bar\lambda^{-}_a\bar
\theta^{+}_{\dot\alpha}\bar\theta^{-\dot\alpha}V^{--}\right],
\end{split}
\label{2.41}
\end{equation}
where $\lambda^{-a}=\lambda^{ia}u^{ -}_i$,
$\bar\lambda^{-a}=\bar\lambda^{ia}u^{ -}_i$,
$\overline{\lambda_{ia}}=\bar \lambda^{ia}$, $\lambda^{ia}$ are the
commuting dimensionless complex parameters. These transformations extend the
$R$-symmetry group from $SU(2)_R\times SU(2)_{\rm PG}$  to  $SU(4)$. The direct check shows that the action (\ref{2.40}) is off-shell invariant under
the transformations (\ref{2.41}). The form of (\ref{2.41}) is almost uniquely specified by
the dimensionality and analyticity reasonings, together with requiring both sides to have the same harmonic $U(1)$ charges.
To avoid a possible confusion, we point out that the superfields in (\ref{2.41}) are not subject to any on-shell conditions
which should be taken into account only when inspecting the closure properties of these transformations (see below).

Further in this section we consider  the case of abelian gauge
group, since the effective action we will deal with depends only on the superfields of the abelian $U(1)$ gauge multiplet.
The equations of motion implied by the action (\ref{2.40})
read
\begin{equation}
D^{++}q^+_a=0,\qquad (D^{+})^2 W=0,\qquad (\bar D^{+})^2{\bar
W}=0.
\label{2.44}
\end{equation}
In addition,  the hypermultiplet $q^+_a$  obeys the off-shell analyticity
constraints
\begin{equation}
D^{+}_{\alpha}q^{+}_{a}=0, \qquad \bar D^{+}_{\dot \alpha}q^{+}_{a}=0.
\end{equation}
The superfield strengths $W$, $\bar W$ are chiral and antichiral
\begin{equation}
\bar D^{\pm}_{\dot{\alpha}}W=0\,, \qquad D^{\pm}_{ \alpha}\bar W=0\,,
\label{01.11}
\end{equation}
and they satisfy the off-shell constraints
\begin{equation}
D^{\pm\pm}W=0\,, \qquad D^{\pm\pm}\bar W=0\,,
\label{01.12}
\end{equation}
which follow from the harmonic flatness condition (\ref{0004}) and the analyticity of $V^{++}$.

When superfields  $ W$, ${\bar W}$ and $q^{+}_a$ obey the on-shell
constraints (\ref{2.44}), the transformations of hidden
$\mathcal{N}=2$ supersymmetry (\ref{00.5}) are simplified to
\begin{equation}
\begin{split}
&\delta  W=\frac{1}{2}\bar\epsilon^{\dot \alpha a}\bar D^{-}_{\dot\alpha} q^+_a,
\qquad \delta {\bar W}=\frac{1}{2}\epsilon^{ \alpha a} D^{-}_\alpha q^+_a,\\
& \delta q^{+}_a=\frac{1}{4}\left(\epsilon^{\alpha}_aD^{+}_\alpha
W+\bar \epsilon^{\dot \alpha}_a\bar D^{+}_{\dot{\alpha}}{\bar
W}\right), \qquad \delta
q^{-}_a=\frac{1}{4}\left(\epsilon^{\alpha}_aD^{-}_\alpha  W+\bar
\epsilon^{\dot \alpha}_a\bar D^{-}_{\dot{\alpha}}{\bar W}\right),
\end{split}
\label{2.45}
\end{equation}
where $q^{-a}=D^{--}q^{+a}$. In this case the $R$-symmetry transformations (\ref{2.41}) are also simplified:
    \begin{equation}
    \begin{split}
    &\delta q^{+}_a=\frac{1}{4}\big(\lambda^{+}_a  W-\lambda^{+}_a\theta^{-\alpha}D^{+}_\alpha W
    +\lambda^{-}_a\theta^{+\alpha}D^{+}_\alpha W+\bar\lambda^{+}_a{\bar W}-
    \bar \lambda^{+}_a\bar\theta^{-\dot\alpha}\bar D^{+}_{\dot\alpha}{\bar W} + \bar \lambda^{-}_a\bar\theta^{+\dot\alpha}\bar D^{+}_{\dot\alpha} {\bar W} \big),\\
    &\delta{\bar W}=\frac{1}{2}\left(\lambda^{-a}q^{+}_{a}-\lambda^{+a}q^{-}_{a}-\lambda^{-a}\theta^{+\alpha}D^{+}_{\alpha} q^{-}_a+\lambda^{+a}\theta^{-\alpha}D^{+}_{\alpha} q^{-}_a\right),\\
    &\delta   W=\frac{1}{2}\left(\bar\lambda^{-a}q^{+}_{a}-\bar\lambda^{+a}q^{-}_{a}-\bar\lambda^{-a}\bar\theta^{+\dot\alpha}\bar D^{+}_{\dot\alpha} q^{-}_a+\bar\lambda^{+a}\bar\theta^{-\dot\alpha}\bar D^{+}_{\dot\alpha} q^{-}_a\right).
    \end{split}
    \label{3333}
    \end{equation}

One may verify that the commutators of the $R$-symmetry transformations
(\ref{3333}) with the manifest and hidden supersymmetry transformations give the
consistent results. The variation of general superfield under
the manifest supersymmetry reads
    \begin{equation}
    \hat \delta \Phi=-\epsilon^{ +\alpha}\frac{\partial\Phi}{\partial \theta^{+\alpha}}
    -\epsilon^{ -\alpha}\frac{\partial\Phi}{\partial \theta^{-\alpha}}
    -\bar\epsilon^{ +}_{\dot\alpha}\frac{\partial\Phi}{\partial\bar \theta^{+}_{\dot\alpha}}
    -\bar\epsilon^{ -}_{\dot\alpha}\frac{\partial\Phi}{\partial \bar\theta^{-}_{\dot\alpha}}
    +2i(\epsilon^{ -\alpha}\bar \theta^{ + \dot \alpha}+\theta^{+\alpha}\bar \epsilon^{ - \dot \alpha}){\partial_{\alpha\dot{\alpha}} \Phi}.
    \label{01.17}
    \end{equation}

    Let us first consider the commutators of the hidden supersymmetry transformations (\ref{2.45})
    with the $R$-symmetry transformations (\ref{3333}). One can show, by a direct computation, that
    \begin{equation}
    \begin{split}
    &\begin{split}
    (\delta_\lambda\delta_\epsilon-\delta_\epsilon\delta_\lambda)q^{+a}=\frac{1}{8}\Big[
    -\bar\lambda^{+}_c\epsilon^{\alpha c}\frac{\partial q^{+a}}{\partial \theta^{+\alpha}}&
    -\lambda^{+}_c\bar\epsilon^{\dot\alpha c}\frac{\partial q^{+a}}{\partial \bar \theta^{+ \dot \alpha}}\\
    &+2i(\bar\lambda^{-}_c\epsilon^{\alpha c}\bar\theta^{+\dot\beta}+\theta^{+\alpha}\lambda^{-}_c\bar \epsilon^{\dot \beta c})
    \partial_{\alpha \dot\beta}q^{+a}\Big]=\frac{1}{8} \hat\delta_{\bar \lambda^{+}_c\epsilon^{\alpha c}}q^{+a},
    \end{split}\\
    &\begin{split}
    (\delta_\lambda\delta_\epsilon-\delta_\epsilon\delta_\lambda)W=\frac{1}{8}\Big[-\bar\lambda^{+}_c\epsilon^{\alpha c}\frac{\partial W}{\partial \theta^{+\alpha}}
    &-\bar\lambda^{-}_c\epsilon^{\alpha c}\frac{\partial W}{\partial \theta^{-\alpha}}-\lambda^{-}_c\bar\epsilon^{\dot\alpha c}\frac{\partial W}{\partial \bar \theta^{- \dot \alpha}}\\
    &+2i(\bar\lambda^{-}_c\epsilon^{\alpha c}\bar\theta^{+\dot\beta}+\theta^{+\alpha}\lambda^{-}_c\bar \epsilon^{\dot \beta c})\partial_{\alpha \dot\beta}W\Big]=\frac{1}{8}\hat \delta_{\bar \lambda^{+}_c\epsilon^{\alpha c}}W,
    \end{split}\\
    &\begin{split}
    (\delta_\lambda\delta_\epsilon-\delta_\epsilon\delta_\lambda)\bar
    W=\frac{1}{8}\Big[
    -\bar\lambda^{-}_c\epsilon^{\alpha c}\frac{\partial \bar W}{\partial \theta^{-\alpha}}&
    -\lambda^{-}_c\bar\epsilon^{\dot\alpha c}\frac{\partial \bar W}{\partial \bar \theta^{- \dot \alpha}}
    -\lambda^{+}_c\bar\epsilon^{\dot\alpha c}\frac{\partial \bar W}{\partial \bar \theta^{+ \dot \alpha}}\\
    &+2i(\bar\lambda^{-}_c\epsilon^{\alpha c}\bar\theta^{+\dot\beta}+\theta^{+\alpha}\lambda^{-}_c\bar \epsilon^{\dot \beta c})\partial_{\alpha \dot\beta}\bar W\Big]=\frac{1}{8}\hat \delta_{\bar \lambda^{+}_c\epsilon^{\alpha c}}\bar W.
    \end{split}
    \end{split}
    \end{equation}
Hence, the on-shell commutator of the hidden supersymmetry transformations (\ref{2.45}) with the $R$-symmetry
ones (\ref{3333}) gives the manifest supersymmetry (\ref{01.17}), with the bracket parameter
$\bar \lambda^{+}_c\epsilon^{\alpha c}$, in agreement with ${\cal N}=4$ supersymmetry algebra.

Let us now evaluate the commutators of  the $R$-symmetry transformations (\ref{3333}) with
the manifest supersymmetry transformations (\ref{01.17}). We obtain
    \begin{equation}
    \begin{split}
    &(\delta_\lambda\hat\delta_\epsilon-\hat\delta_\epsilon\delta_\lambda)q^{+}_{a}=-\frac{1}{4}\left(\epsilon^{\alpha i}\lambda_{ia}D^{+}_\alpha
    W+\bar \epsilon^{\dot \alpha i}\bar \lambda_{ia}\bar D^{+}_{\dot{\alpha}}{\bar W}\right)=- \delta_{\lambda_{ia}\epsilon^{\alpha i}}q^{+}_{a},\\
    &(\delta_\lambda\hat\delta_\epsilon-\hat\delta_\epsilon\delta_\lambda)W=-\frac{1}{2}\bar \epsilon^{\dot \alpha i}\bar \lambda^a_{i}\bar D^{-}_{\dot\alpha} q^+_a
    =- \delta_{\lambda_{ia}\epsilon^{\alpha i}}W,\\
    &(\delta_\lambda\hat\delta_\epsilon-\hat\delta_\epsilon\delta_\lambda)\bar W=-\frac{1}{2}\epsilon^{\alpha i}\lambda^a_{i} D^{-}_\alpha q^+_a
    =- \delta_{\lambda_{ia}\epsilon^{\alpha i}}\bar W.
    \end{split}
    \end{equation}
So the on-shell commutator of  the $R$-symmetry transformations (\ref{3333}) with the manifest
supersymmetry transformations (\ref{01.17}) yields those of hidden supersymmetry (\ref{2.45}),
with the bracket parameter $(\epsilon^\alpha_{a}))_{\rm br} = \lambda_{ia}\epsilon^{\alpha i}$.

    Finally, we consider the commutator of the  $R$-symmetry transformations with itself. We have
    \begin{equation}
    \begin{split}
    (\delta_{\lambda_1}\delta_{\lambda_2}-\delta_{\lambda_2}\delta_{\lambda_1})q^{+}_a=&\frac{1}{8}
    \left[\lambda^{b}_{\text{(PG)}a}q^{+}_b+\frac{1}{2}\lambda\bar \theta^{+\dot\alpha}\frac{\partial}{\partial \bar \theta^{+\dot \alpha}}q^{+}_a
    +\frac{1}{2}\bar\lambda \theta^{+\alpha}\frac{\partial}{\partial  \theta^{+ \alpha}}q^{+}_a\right]\\
    &-\frac{1}{8}\lambda^{i}_j\left(u^{+}_i\frac{\partial}{\partial u^{+j}}+u^{-}_i\frac{\partial}{\partial u^{-j}}\right)q^{+}_a,
    \\
    (\delta_{\lambda_1}\delta_{\lambda_2}-\delta_{\lambda_2}\delta_{\lambda_1})\bar W=
    &-\frac{1}{8}\lambda^{i}_j\left(u^{+}_i\frac{\partial}{\partial u^{+j}}+u^{-}_i\frac{\partial}{\partial u^{-j}}\right)\bar W \\
    &+\frac{1}{8}\bigg[\frac{1}{2}\lambda \bar \theta^{-\dot \alpha}\frac{\partial}{\partial\bar\theta^{-\dot{\alpha}}}\bar W
    + \frac{1}{2}\lambda \bar \theta^{+\dot \alpha}\frac{\partial}{\partial\bar\theta^{+\dot{\alpha}}}\bar W
    + \frac{1}{2}\bar \lambda  \theta^{+ \alpha}\frac{\partial}{\partial\theta^{+{\alpha}}}\bar W-\lambda \bar W
    \bigg],
    \label{2.19}
    \end{split}
    \end{equation}
    where
    \begin{equation}
    \lambda^{ij}=\lambda^{(ia}_2\bar \lambda^{j)}_{1a}-\lambda^{(ia}_1\bar \lambda^{j)}_{2a}, \qquad \lambda=\lambda^{ia}_2\bar \lambda_{1ia}-\lambda^{ia}_1\bar \lambda_{2ia},
    \qquad
    \lambda^{ab}_{\text{(PG)}}=\lambda^{i(a}_{2}\bar \lambda^{b)}_{1i}-\lambda^{i(a}_{1}\bar \lambda^{b)}_{2i}.
    \end{equation}
Here, the parameters $\lambda^{ij}$ correspond to $SU(2)_R$ transformations, the parameter  $\lambda$ ($\bar \lambda=-\lambda$)
corresponds to the additional  $U(1)_{R}$ symmetry
and $\lambda^{ab}_{\text{(PG)}}$ are associated with the $SU(2)_\text{PG}$ symmetry commuting with both ${\cal N}=2$
supersymmetry and $U(2)_R$ symmetry.

Thus the on-shell closure of the  implicit $R$-symmetry transformations  yields the linear $U(2)_R$ and $SU(2)_{\text{PG}}$
transformations, once again in agreement with the action of the coset $SU(4)_{R}/[U(2)_R\times SU(2)_{\text{PG}}]$ part of
the full automorphism symmetry $SU(4)_{R}$ on ${\cal N}=4$ supersymmetry algebra.

    Note that the calculation of the brackets (\ref{2.19}) is not as straightforward as that of the previous Lie brackets. Some details of it
    are collected in Appendix.

\subsection{Effective action}

The leading low-energy term in the effective action of $ \mathcal {N} =
2 $ SYM theory in $ \mathcal {N} = 2 $ superspace has the form (see,
e.g., the reviews \cite{litlink 23, litlink 22})
\begin{equation}
{\Gamma}_0 = \int d^{12}zdu\,\mathcal{H}( W, {\bar W}), \qquad
\mathcal{H}( W, {\bar W})=c\ \text{ln}\left(\frac{
W}{\Lambda}\right)\text{ln}\left(\frac{{\bar W}}{\Lambda}\right),
\label{2.46}
\end{equation}
where $\Lambda$ is an arbitrary scale\footnote{In fact, the action does not depend on $\Lambda$ in virtue
of the (anti)chirality of $(\bar{W})W$.}.

The complete $\mathcal{N}=4$ SYM effective action is an extension of
the effective action (\ref{2.46}) by hypermultiplet-dependent terms.
It was first found in \cite{litlink 8}:
    \begin{gather}
    \Gamma=\int d^{12}zdu\ \left[c\ \text{ln}\left(\frac{ W}{\Lambda}\right)
    \text{ln}\left(\frac{{\bar W}}{\Lambda}\right)+\mathcal{L}\left(-2\frac{q^{+a}q^{-}_{a}}{ W{\bar W}}
    \right)\right],\label{2.50}
    \end{gather}
    where
     \begin{gather}
    \mathcal{L}(Z)=c\sum\limits_{n=1}^{\infty}\frac{Z^n}{n^2(n+1)}=c\left[(Z-1)\frac{\ln (1-Z)}{Z}
    +\text{Li}_2(Z)-1\right]
    \end{gather}
and $\text{Li}_2(Z)$ is the Euler dilogarithm function. The part
containing $ q^{+a} $ is fixed by the requirement that the effective
action $\Gamma $ be invariant under both manifest off-shell $ \mathcal {N} =
2$ supersymmetry and hidden on-shell $ \mathcal {N} = 2$
supersymmetry. As a result, the effective action (\ref{2.50}) is an invariant of
$\mathcal {N} = 4$ supersymmetry and depends on all fields of $ \mathcal {N} = 4$
gauge multiplet.

Now we will show that the effective action (\ref{2.50}) can be
alternatively derived from (\ref{2.46})  by imposing the requirement of $R$-symmetry
instead of invariance under the hidden $ \mathcal {N} = 2$
supersymmetry. To this end, let us consider the variation of
    (\ref{2.46})  under the transformations (\ref{3333}). Based on the reality reasonings, it is enough to concentrate only on that part of the transformations which involves
    the parameter $\lambda^{ia}$,  neglecting the part containing $\bar\lambda^{ia}$,
    \begin{equation}
    \begin{split}
    &\delta\int d^{12}zdu\ c\ \text{ln}\left(\frac{ W}{\Lambda}\right)\text{ln}\left(\frac{{\bar W}}{\Lambda}\right)\\
    &=\frac{c}{2} \int d^{12}zdu \ \text{ln}\left(\frac{ W}{\Lambda}\right)\frac{2\lambda^{-a}q^{+}_a
    +\lambda^{-a}\theta^{+\alpha}D^{-}_\alpha q^{+}_a-\lambda^{+a}\theta^{-\alpha}D^{-}_\alpha q^{+}_a}{{\bar W}} \\
    &=\frac{c}{2} \int d^{12}zdu  \frac{-\lambda^{-a}\theta^{+\alpha}D^{-}_\alpha  W q^{+}_a+\lambda^{+a}\theta^{-\alpha}D^{-}_\alpha  W  q^{+}_a}{ W{\bar W}}.
    \end{split}
    \label{16777}
    \end{equation}
     Due to the property
    that  $\int d^{12}zdu \frac{q^{+a}\lambda^{-}_a}{{\bar W}}=0$, the variation (\ref{16777}) can be canceled by adding
    the new term to ${\cal H}$:
    \begin{equation}
    \mathcal{ L}_1=-c\ \frac{q^{+a}q^-_{a}}{W\bar W}.
    \end{equation}
    Evaluating the variation of the sum ${\cal H} + {\cal L}_1$, we obtain
    \begin{equation}
    \begin{split}
    &
    \delta\int d^{12}zdu \Big[ \mathcal{H}({W},{ \bar W})+\mathcal{ L}_1\Big]=
    \frac{c}{2}\int d^{12}zdu \frac{q^{+b}q^{-}_b}{ W{\bar W}^2}
    \big[\lambda^{-a}q^{+}_a-\lambda^{+a}q^{-}_a\\
    &+\left(\lambda^{-a}\theta^{+\alpha}-\lambda^{+a}\theta^{-\alpha}\right)D^{+}_\alpha q^{-}_a\big]
    \\
    &=\frac{c}{2}\int d^{12}zdu  \frac{q^{+b}q^{-}_b}{{W}{\bar W}^2}
    \left[2\lambda^{-a}q^{+}_a+(\lambda^{-a}\theta^{+\alpha}-\lambda^{+a}\theta^{-\alpha})D^{-}_\alpha q^{+}_a\right].
    \end{split}
    \label{2.23}
    \end{equation}
    Consider the last term in some detail
    \begin{equation}
    \begin{split}
    &\frac{c}{2}\int d^{12}zdu  \frac{q^{+b}q^{-}_b}{{W}{\bar W}^2}
    (\lambda^{-a}\theta^{+\alpha}-\lambda^{+a}\theta^{-\alpha})D^{-}_\alpha q^{+}_a\\
    &  =\frac{c}{2}\int d^{12}zdu\bigg[ \frac{
        -2\lambda^{-a} W+(\lambda^{-a}\theta^{+\alpha}- \lambda^{+a}\theta^{-\alpha})D^{-}_\alpha  W }{({W}{\bar W})^2}q^{+}_a(q^{+b}q^{-}_b)\\
    &-(\lambda^{-a}\theta^{+\alpha} -\lambda^{+a}\theta^{-\alpha})\frac{q^{+b}D^{-}_\alpha q^{-}_bq^{+}_a}{ {W}{\bar W}^2}\bigg]
    \\
    &=\frac{c}{2}\int d^{12}zdu\bigg[ \frac{
        -2\lambda^{-a} W+(\lambda^{-a}\theta^{+\alpha}- \lambda^{+a}\theta^{-\alpha})D^{-}_\alpha  W }{({W}{\bar W})^2}q^{+}_a(q^{+b}q^{-}_b)\\
    &-(\lambda^{-a}\theta^{+\alpha} -\lambda^{+a}\theta^{-\alpha})\frac{q^{+b}q^{-}_bD^{-}_\alpha q^{+}_a}{2 {W}{\bar W}^2}\bigg].
    \end{split}
    \end{equation}
    Here we have used various properties of the involved superfields (chirality, analyticity), as well as the integration by parts with respect to the spinor derivative in the second line and cyclic identities for the $SU(2)$ doublet indices in the third line. Observe that the last term
    in the third line equals, modulo the minus sign, the variation we started with. Hence,
    \begin{equation}
    \begin{split}
    &\frac{c}{2}\int d^{12}zdu  \frac{q^{+b}q^{-}_b}{{W}{\bar W}^2}
    (\lambda^{-a}\theta^{+\alpha}-\lambda^{+a}\theta^{-\alpha})D^{-}_\alpha q^{+}_a \\
    &=\frac{c}{3}\int d^{12}zdu\bigg[ \frac{
        -2\lambda^{-a} W+(\lambda^{-a}\theta^{+\alpha}- \lambda^{+a}\theta^{-\alpha})D^{-}_\alpha  W }{({W}{\bar W})^2}q^{+}_a(q^{+b}q^{-}_b)\bigg].
    \end{split}
    \label{2.25}
    \end{equation}
    Plugging (\ref{2.25}) into (\ref{2.23}), we obtain
    \begin{equation}
    \begin{split}
    &\delta\int d^{12}zdu \left[ \mathcal{H}({W},{ \bar W})+\mathcal{ L}_1 \right]\\
    &=
    \frac{c}{3}\int d^{12}zdu  \frac{q^{+b}q^{-}_b}{({W\bar W})^2}
    \left[\lambda^{-a}q^{+}_a W+\left(\lambda^{-a}\theta^{+\alpha}-\lambda^{+a}\theta^{-\alpha}\right)D^{-}_\alpha  W q^{+}_a\right].
    \end{split}
    \end{equation}

    This variation is canceled by adding the new proper term to (\ref {2.46}),
    \begin{equation}
    \mathcal{ L}_2=\frac{c}{3} \frac{(q^{+a}q^-_{a})^2}{({W\bar W})^2}.
    \end{equation}
    Continuing the iterative process,  one can find that
    \begin{equation}
    \mathcal{ L}_n=\frac{c}{n^2(n+1)} \frac{(-2q^{+a}q^-_{a})^n}{({W\bar W})^n}.
    \end{equation}
    Summing up all $ \mathcal {L} _n $, one recovers the effective action (\ref {2.50}).

    One can directly verify that (\ref{2.50}) is invariant under the $R$-symmetry transformations (\ref{3333}).
    Once again, we limit our attention to terms with the parameter $\lambda^{ia}$:
    \begin{equation}
    \begin{split}
    &\delta \Gamma =c \int d^{12}zdu\bigg[ \frac{\left(\lambda^{+a}\theta^{-\alpha}-\lambda^{-a}\theta^{+\alpha}\right)D^{-}_\alpha  W q^{+}_a}{2 W{\bar W}}\\
    &-\mathcal{ L}'(Z)\frac{q^{+a}\left[\lambda^{-}_a  W
    +(\lambda^{-}_a\theta^{+\alpha}-\lambda^{+}_a\theta^{-\alpha})D^{-}_\alpha W\right]}{W\bar W}
    \\
    &-Z\mathcal{ L}'(Z)\frac{
    \left[2\lambda^{-a}q^{+}_a+(\lambda^{-a}\theta^{+\alpha}-\lambda^{+a}\theta^{-\alpha})D^{-}_\alpha q^{+}_a\right]}{2{\bar W}}\bigg].
    \end{split}
    \end{equation}
    This expression can be simplified, using the identity
    \begin{equation}
    \begin{split}
    &\int d^{12}zdu \frac{Z\mathcal{ L}'(Z)}{\bar W} \left[2\lambda^{-a}q^{+}_{a}+(\lambda^{-a}\theta^{+\alpha}-\lambda^{+a}\theta^{- \alpha})D^{-}_{ \alpha} q^{+}_a\right]\\
    &=\int d^{12}zdu\ [2\mathcal{ L}'(Z)-1]\bigg[\frac{\lambda^{-a}q^{+}_a}{\bar W}+(\lambda^{-a}\theta^{+\alpha}-\lambda^{+a}\theta^{- \alpha}) \frac{q^{+}_aD^{-}_{ \alpha}W}{W\bar W}\bigg],
    \end{split}
    \end{equation}
    which is deduced by integrating by parts with respect to the spinor derivative and applying to
    the definition of the function $\mathcal{ L}(Z)$ (\ref{2.50}).

    Thus we obtain
\begin{equation}
\begin{split}
&\delta \Gamma =c \int d^{12}zdu\bigg[ \frac{\left(\lambda^{+a}\theta^{-\alpha}-\lambda^{-a}\theta^{+\alpha}\right)D^{-}_\alpha  W q^{+}_a}{2 W{\bar W}}\\
&-\mathcal{ L}'(Z)\frac{q^{+a}\left[\lambda^{-}_a  W
    +(\lambda^{-}_a\theta^{+\alpha}-\lambda^{+}_a\theta^{-\alpha})D^{-}_\alpha W\right]}{W\bar W}\bigg]
\\
&-\frac{1}{2} \int d^{12}zdu\ [2\mathcal{ L}'(Z)-1]\bigg[\frac{\lambda^{-a}q^{+}_a}{\bar W}+(\lambda^{-a}\theta^{+\alpha}-\lambda^{+a}\theta^{- \alpha}) \frac{q^{+}_aD^{-}_{ \alpha}W}{W\bar W}\bigg]=0.
\end{split}
\end{equation}

To summarize,   the requirement of invariance under the $R$-symmetry
transformations allowed us to  completely restore the hypermultiplet
dependence of $ \mathcal {N} = 4$ supersymmetric effective action.

\subsection{Superconformal invariance of the effective action}

The effective action (\ref{2.50}) is evidently scale-invariant. In this section we prove that  it
is actually invariant under the total $4D,\, {\cal N}=4$ superconformal group $SU(2,2|4)$.

Due to the structure of $4D,\, \mathcal{ N}=4$ superconformal
algebra and the $R$-symmetry invariance of the effective action it suffices to show only its $4D,\,\mathcal{ N}=2$ superconformal
invariance. Moreover, it is enough to check just invariance under conformal boosts.

Indeed, the commutator of conformal boosts with the manifest ${\cal N}=2$ Poincar\'e supersymmetry yields
the special conformal ${\cal N}=2$ supersymmetry. The ${\cal N}=4$ completion of the latter is contained
in the commutator of conformal boosts with the hidden ${\cal N}=2$ supersymmetry which, as we saw, is obtained by commuting
the hidden $R$-symmetry with ${\cal N}=2$ Poincar\'e supersymmetry.

To prove the superconformal invariance we should use the
transformation rules of the harmonic superspace coordinate, as well as those of the harmonic and spinor derivatives. The
harmonic superspace coordinate transformations under conformal boosts in the analytic basis read
\cite{litlink 1}
    \begin{equation}
    \begin{split}
    &   \delta x^{\alpha \dot \alpha}=x^{\beta\dot{\alpha}}k_{\beta\dot \beta}x^{\alpha\dot \beta},\\
    &\delta\theta^{+\alpha}=\theta^{+\beta}k_{\beta\dot \beta}x^{\alpha\dot{\beta}},\qquad
    \delta\theta^{-\alpha}=\theta^{-\beta}k_{\beta\dot \beta}x^{\alpha\dot{\beta}}-2i(\theta^{-})^2\bar \theta^{+}_{\dot{\beta}}k^{\alpha\dot{\beta}},\\
    &\delta\bar\theta^{+\dot\alpha}=\bar\theta^{+\dot\beta}k_{\beta\dot \beta}x^{\beta\dot{\alpha}},\qquad
    \delta\bar\theta^{-\dot\alpha}=\bar\theta^{-\dot\beta}k_{\beta\dot \beta}x^{\beta\dot{\alpha}}-2i(\bar\theta^{-})^2 \theta^{+}_{{\beta}}k^{\beta\dot{\alpha}},
    \end{split}
    \label{x}
    \end{equation}
    where $k^{\alpha\dot \alpha}$ is the corresponding $4$-vector parameter. The transformation law of the harmonics is
\cite{litlink 1}
    \begin{equation}
    \begin{split}
    &\delta u^{+k}=\Lambda^{++}u^{-}_k, \qquad \delta u^{-}_k=0, \qquad
    \Lambda^{++}=4i\theta^{+\alpha} k_{\alpha\dot\alpha}\bar\theta^{+\dot{\alpha}},\\
    &D^{--} \Lambda^{++}=4i\theta^{-\alpha} k_{\alpha\dot\alpha}\bar\theta^{+\dot{\alpha}}+
    4i\theta^{+\alpha} k_{\alpha\dot\alpha}\bar\theta^{-\dot{\alpha}}.
    \end{split}
    \label{u}
    \end{equation}
Next, let us write the superconformal transformations of the harmonic
and spinor derivatives
    \begin{equation}
    \delta D^{++}=-\Lambda^{++}D^{0}, \qquad \delta D^{--}=-(D^{--}\Lambda^{++})D^{--}, \qquad \delta D^{+}_{\alpha}=-D^{+}_\alpha(\delta \theta^{-\beta})D^{+}_\beta.
    \label{1.22}
    \end{equation}
    Using these transformation rules it is easy to establish the transformation of the superspace
    integration measure $dZ=d^{12}zdu=d^4xd^4\theta^{+}d^4\theta^{-}du$,
    \begin{equation}
    \begin{split}
    &\delta dZ
    =\left(\partial_{\alpha{\dot \alpha}} x^{\alpha{\dot \alpha}}+\partial^{--}\Lambda^{++}-\partial_{+\alpha}\theta^{+\alpha}-\partial_{-\alpha}\theta^{-\alpha}-\bar\partial_{+\dot\alpha}\bar\theta^{+\dot\alpha}-\bar\partial_{+\dot\alpha}\bar\theta^{+\alpha}  \right)dZ=\Lambda dZ,\\
    &\Lambda=4i(\theta^-_\alpha\bar\theta^{+}_{\dot{\alpha}}-\bar\theta^{-}_{\dot{\alpha}}\theta^{+}_{\alpha})k^{\alpha\dot{\alpha}}.
    \end{split}
    \end{equation}

Let us now consider the transformations of the gauge potentials
$V^{\pm\pm}$. Taking into account the relations (\ref{x}),
(\ref{u}), (\ref{1.22}) one deduces
\begin{equation}
    \delta V^{++}=0, \qquad \delta V^{--}=-(D^{--}\Lambda^{++})V^{--}.
    \label{1.24}
    \end{equation}
    Using the transformation rules (\ref{1.22}) and (\ref{1.24}), it is easy to obtain the transformation law of the
    superfield strengths $W$, $\bar W$
    \begin{equation}
    \delta W=-k_{\alpha {\dot \alpha}}(x^{\alpha {\dot \alpha}}+4i\theta^{-\alpha}\bar \theta^{+\dot{\alpha}})W, \qquad \delta \bar W=-k_{\alpha {\dot \alpha}}(x^{\alpha {\dot \alpha}}+4i\theta^{+\alpha}\bar \theta^{-\dot{\alpha}})\bar W.
    \end{equation}
The transformation of the hypermultiplet $q^{\pm a}$ under the conformal boosts reads \cite{litlink 1}
    \begin{equation}
    \delta q^{+a}=-k_{\alpha {\dot \alpha}}x^{\alpha{\dot \alpha}}q^{+a}, \qquad \delta q^{-a}=\delta(D^{--}q^{+a})=-k_{\alpha {\dot \alpha}}x^{\alpha{\dot \alpha}}q^{-a}-4i(\theta^-_\alpha\bar\theta^{+}_{\dot{\alpha}}-\bar\theta^{-}_{\dot{\alpha}}\theta^{+}_{\alpha})k^{\alpha\dot{\alpha}}q^{-a}.
    \end{equation}

Now we are prepared to show  the conformal invariance of the effective action (\ref{2.50}).

Let us first show invariance of the logarithmic term
    \begin{equation}
    \begin{split}
    &\delta \int d^{12}zdu c\ \text{ln}\left(\frac{ W}{\Lambda}\right)\text{ln}\left(\frac{{\bar W}}{\Lambda}\right)
    =c \int d^{12}zdu  \Big[ 4i(\theta^-_\alpha\bar\theta^{+}_{\dot{\alpha}}-\bar\theta^{-}_{\dot{\alpha}}\theta^{+}_{\alpha})k^{\alpha\dot{\alpha}}
    \text{ln}\left(\frac{ W}{\Lambda}\right)\text{ln}\left(\frac{{\bar W}}{\Lambda}\right)\\
    &-k_{\alpha {\dot \alpha}}(x^{\alpha {\dot \alpha}}+4i\theta^{-\alpha}\bar \theta^{+\dot{\alpha}})\text{ln}\left(\frac{{\bar W}}{\Lambda}\right)
    -k_{\alpha {\dot \alpha}}(x^{\alpha {\dot \alpha}}+4i\theta^{+\alpha}\bar \theta^{-\dot{\alpha}})
    \text{ln}\left(\frac{{ W}}{\Lambda}\right)\Big]\\
    &= c\ \int d^{12}zdu  \Big[ -4i(\theta^+_\alpha\bar\theta^{-}_{\dot{\alpha}}+\bar\theta^{-}_{\dot{\alpha}}\theta^{+}_{\alpha})k^{\alpha\dot{\alpha}}  \text{ln}\left(\frac{ W}{\Lambda}\right)\text{ln}\left(\frac{{\bar W}}{\Lambda}\right)\Big]=0,
    \end{split}
    \end{equation}
where we made use of some properties of $W$ and $\bar W$, eqs.  (\ref{01.11}), (\ref{01.12}).
Then we check invariance of the generic term in the power expansion of the function $\mathcal{L}(z)$
    \begin{equation}
    \begin{split}
    &\delta \int d^{12}zdu \left(\frac{q^{+a}q^{-}_{a}}{W\bar W}\right)^n=
    n \int d^{12}zdu\left(\frac{q^{+a}q^{-}_{a}}{W\bar W}\right)^n\Big[ \frac{4i}{n}(\theta^-_\alpha\bar\theta^{+}_{\dot{\alpha}}
    -\bar\theta^{-}_{\dot{\alpha}}\theta^{+}_{\alpha})k^{\alpha\dot{\alpha}}\\
    &-k_{\alpha {\dot \alpha}}x^{\alpha{\dot \alpha}} -k_{\alpha {\dot \alpha}}x^{\alpha{\dot \alpha}}
    -4i(\theta^-_\alpha\bar\theta^{+}_{\dot{\alpha}}-\bar\theta^{-}_{\dot{\alpha}}\theta^{+}_{\alpha})k^{\alpha\dot{\alpha}}\\
    &+k_{\alpha {\dot \alpha}}(x^{\alpha {\dot \alpha}}+4i\theta^{-\alpha}\bar \theta^{+\dot{\alpha}})+
    k_{\alpha {\dot \alpha}}(x^{\alpha {\dot \alpha}}+4i\theta^{+\alpha}\bar \theta^{-\dot{\alpha}})\Big]=0.
    \end{split}
    \end{equation}
Here we used, once again, the  conditions (\ref{01.11}) and (\ref{01.12}),
as well as the equations of motion (\ref{2.44}).

    So we have proved that the effective action (\ref{2.50}) is superconformally invariant on shell. Note that the original
    ``microscopic'' action is invariant under the conformal boosts and hence under the whole ${\cal N}=4$
    superconformal group off shell, without any use of the equations of motion. The latter, like in the case of hidden ${\cal N}=4$ supersymmetry and $R$-symmetry,
    are of need only when checking the correct closure of all these transformations. On the other hand, the effective action reveals the invariance under the
    hidden ${\cal N}=4$ supersymmetry and $R$-symmetry only on shell, so it is quite natural that the same is also valid for ${\cal N}=4$ superconformal symmetry.

    \section{$5D, \,\mathcal{ N}=2$  SYM theory}
In this section, we introduce the $R$-symmetry transformations for $5D,\, \mathcal{N} = 2$ SYM
theory with the gauge group $SU(2)$ and construct the complete low-energy effective action by
requiring invariance under these $R$-symmetry transformations.
We use the notations and conventions of Refs. \cite{litlink 1,litlink 11}. The relevant harmonic superspace formalism to large extent is
similar to its $4D,\, {\cal N}=2$ prototype.

\subsection{Classical action}
    The $\mathcal{ N}=2$ gauge multiplet in  $5D$, $\mathcal{ N}=1$ harmonic superspace consists of $\mathcal{ N}=1$ gauge multiplet represented by the analytic superfield
    $V^{++}$ and the hypermultiplet $q^{+a}$.
    The $\mathcal{ N}=1$ gauge multiplet  classical action reads \cite{litlink 9}
    \begin{equation}
    S^{ \mathcal{N}=1}_{\text{SYM}}=\frac{1}{2g^2}\sum\limits_{n=2}^{\infty}\text{tr}\frac{(-i)^n}{n}\int d^{13}z du_1\dots du_n\frac{V^{++}(z,u_1)\dots V^{++}(z,u_n)}{(u^+_1u^+_2)\dots(u^+_nu^+_1)},
    \label{3.1}
    \end{equation}
    where $g$ is a coupling constant of mass-dimension $-1/2$. The superfield strength is defined in the analytic $\lambda$-frame as
    \begin{equation}
    W=\frac{i}{8}(D^{+})^2V^{--},
    \label{3.2}
    \end{equation}
    where $(D^{+})^2= D^{+\hat \alpha}D^{+}_{\hat \alpha} = \Omega^{\hat \alpha \hat \beta} D^{+}_{\hat \beta}D^{+}_{\hat \alpha}$, $\;\Omega^{\hat \alpha \hat \beta}$ is
    the $USp(4)$ invariant skew-symmetric constant ``metric''   and $V^{--}$ is a non-analytic gauge potential
    related to $V^{++}$ by the harmonic flatness condition
    \begin{equation}
    D^{++}V^{--}-D^{--}V^{++}+i[V^{++}, V^{--}]=0.
    \label{3.3}
    \end{equation}

    The classical action of the hypermultiplet $q^{+a}=(q^{+}, -\tilde q^{+})$ in the adjoint representation of the gauge
    group is written as  \cite{litlink 10}
    \begin{equation}
    S_q=\frac{1}{2g^2}\text{tr}\int d\zeta^{-4}q^{+}_a\nabla^{++}q^{+a}=
    \frac{1}{2g^2}\text{tr}\int d\zeta^{-4}q^{+}_a\left(D^{++}q^{+a}+i[V^{++},q^{+a}]\right),
    \label{3.4}
    \end{equation}
    where $d\zeta^{-4}$ is the analytic superspace integration measure. In addition, this action is invariant under
    $SU(2)_{{\rm PG}}$ symmetry which transforms $q^{+a}$ as a doublet.

    The action of $\mathcal{ N} = 2$ gauge multiplet in $5D, \mathcal N = 1$ harmonic superspace is
    just the sum of (\ref{3.1}) and (\ref{3.4}),
    \begin{equation}
    S^{ \mathcal{N}=2}=S^{ \mathcal{N}=1}_{\text{SYM}}+S_q.
    \label{1.0}
    \end{equation}
    The action is invariant under the implicit $\mathcal{ N}=1$ supersymmetry completing the manifest $\mathcal{ N}=1$ supersymmetry
    to the total  $\mathcal{ N}=2$ supersymmetry
    \begin{equation}
    \delta q^{+}_a=-\frac{1}{2}(D^{+})^4\left[\epsilon_{a\hat{\alpha}}\theta^{-\hat \alpha}V^{--}\right], \qquad \delta V^{++}=\epsilon^a_{\hat \alpha} \theta^{+\hat \alpha}q^{+}_a,
    \label{3.6}
    \end{equation}
    where $\epsilon^{a}_{a\hat \alpha}$ is the relevant anticommuting parameter and
    \begin{equation}
    (D^+)^4=-\frac{1}{32}(D^{+})^2(D^{+})^2.
    \end{equation}

\subsection{$R$-symmetry transformations}
We define the $R$-symmetry transformations in $5D, \,\mathcal{ N}=1$ harmonic superspace as
\begin{equation}
\delta q^{+a}=-\frac{i}{4}(D^{+})^4 \left[\lambda^{+a} (\theta^{- })^2 V^{--}
-2\lambda^{-a}\theta^{+\hat \alpha}\theta^{-}_{\hat \alpha}V^{--}\right],
\qquad \delta V^{++}= -i\lambda^{-a}(\theta^{+})^2q^{+}_a,
\label{3.8}
\end{equation}
where $\lambda^{ia}$ is the relevant commuting parameter
($\overline{\lambda^{ia}}= \lambda_{ia}$,
$\lambda^{+a}=\lambda^{ia}u^{+}_i$). These transformations extend
the $R$-symmetry group from $SU(2)_R\times SU(2)_{\rm PG}$ to
$SO(5)$. The direct check shows that the action (\ref{1.0}) is
invariant off shell under the transformations (\ref{3.8}).

Further in this section we consider the case of abelian gauge group. The action (\ref{1.0})
yields the equations of motion
\begin{equation}
(D^{+})^2 W=0, \qquad D^{++}q^{+}_a=0.
\label{3.9}
\end{equation}
In addition, the superfield strength  $W$ satisfies the off-shell constraints
\begin{equation}
D^{++}W=0,\qquad D^{--}W=0\,,
\label{3.10}
\end{equation}
which follow from the harmonic flatness condition (\ref{3.3}) and the analyticity of $V^{++}$.

On the shell of the  equations of motion (\ref{3.9})  the transformations of the hidden supersymmetry (\ref{3.6}) are reduced to
\begin{equation}
\delta q^{\pm }_a=\frac{i}{2}\epsilon^{\hat{\alpha}}_a(D^{\pm}_{\hat{\alpha}}W), \qquad
\delta W= -\frac{i}{8}\epsilon^a_{\hat{\alpha}}D^{-\hat{\alpha}}q^{+}_a
+\frac{i}{8}\epsilon^a_{\hat{\alpha}}D^{+\hat{\alpha}}q^{-}_a,
\label{1.11}
\end{equation}
where $q^{-a}=D^{--}q^{+a}$.
The $R$-symmetry transformations (\ref{3.8}) take the form
\begin{gather}
\begin{split}
&\delta q^{+}_a=-\frac{1}{2}\left(\lambda^{+}_a  W-\lambda^{+}_a\theta^{-\hat\alpha}D^{+}_{\hat \alpha}  W
+\lambda^{-}_a\theta^{+\hat \alpha}D^{+}_{\hat \alpha} W \right),\\
&\delta W=\frac{1}{4}\left(2\lambda^{+a}q^{-}_{a}-2\lambda^{-a}q^{+}_{a}+\lambda^{-a}\theta^{+\hat\alpha}D^{+}_{\hat \alpha} q^{-}_a-\lambda^{+a}\theta^{-\hat \alpha}D^{+}_{\hat \alpha} q^{-}_a\right).
\end{split}
\label{1.12}
\end{gather}

Now we can consider the commutator of supersymmetry transformations with those of $R$-symmetry.
The variation of general superfield under the manifest supersymmetry reads
    \begin{equation}
    \hat\delta \Phi=-\epsilon^{ +\hat \alpha}\frac{\partial\Phi}{\partial \theta^{+\hat\alpha}}
    -\epsilon^{\hat -\hat \alpha}\frac{\partial\Phi}{\partial \theta^{-\hat \alpha}}
    -2i\epsilon^{-\hat \alpha}\theta^{+\hat\beta}\partial_{\hat\alpha\hat \beta}\Phi,
    \end{equation}
    where $\epsilon^{ \pm}_{\hat \alpha}=\epsilon^{a}_{\hat \alpha} u^{ \pm}_a$ are the relevant anticommuting parameters.

Let us first consider the commutators of the hidden supersymmetry transformations  (\ref{1.11}) with the $R$-symmetry
transformations (\ref{1.12}). One can show that
    \begin{equation}
    \begin{split}
    &(\delta_\lambda\delta_\epsilon-\delta_\epsilon\delta_\lambda)q^{+a}=\frac{i}{8}\left(
    -\lambda^{-}_c\epsilon^{\hat\alpha c}\frac{\partial q^{+a}}{\partial \theta^{-\hat\alpha}}
    -2i\lambda^{-}_c\epsilon^{\hat\alpha c}\theta^{+\hat\beta}\partial_{\hat\alpha\hat \beta}q^{+a}\right)=\frac{i}{8}\hat \delta_{\lambda^{i}_c\epsilon^{\hat\alpha c}}q^{+a},\\
    &(\delta_\lambda\delta_\epsilon-\delta_\epsilon\delta_\lambda)W=\frac{i}{8}\left(
    -\lambda^{+}_c\epsilon^{\hat\alpha c}\frac{\partial W}{\partial \theta^{+\hat\alpha}}
    -\lambda^{-}_c\epsilon^{\hat\alpha c}\frac{\partial W}{\partial \theta^{-\hat\alpha}}
    -2i\lambda^{-}_c\epsilon^{\hat\alpha c}\theta^{+\hat\beta}\partial_{\hat\alpha\hat \beta}W\right)=\frac{i}{8}\hat \delta_{\lambda^{i}_c\epsilon^{\hat\alpha c}}W.
    \end{split}
    \end{equation}
Hence, the on-shell commutator of the hidden supersymmetry transformations with the $R$-symmetry transformations
gives the manifest supersymmetry with the bracket parameter $\lambda^{i}_c\epsilon^{\hat\alpha c}$, as expected.

Let us now consider the commutators of  the $R$-symmetry transformations (\ref{1.12}) with the
manifest supersymmetry transformations (\ref{1.13}). They are given by
    \begin{equation}
    \begin{split}
    &(\delta_\lambda\hat\delta_\epsilon-\hat\delta_\epsilon\delta_\lambda)q^{+a}=-\frac{1}{2}\epsilon^{\hat\alpha}_i\lambda^{ia}D^{+}_{\hat \alpha}W=i\hat \delta_{\lambda^{ia}\epsilon^{\hat\alpha}_i}q^{+a},\\
    &(\delta_\lambda\hat\delta_\epsilon-\hat\delta_\epsilon\delta_\lambda)W=-\frac{1}{4}\epsilon^{\hat\alpha}_i\lambda^{ia}D^{+}_{\hat \alpha}q^{-}_a
    =i\hat \delta_{\lambda^{ia}\epsilon^{\hat\alpha}_i}W.
    \end{split}
    \end{equation}
Thus the on-shell commutator of $R$-symmetry with the manifest supersymmetry yields
the hidden supersymmetry with the bracket  parameter $\lambda^{ia}\epsilon^{\hat\alpha}_i$.

At last, one can consider the commutator of $R$-symmetry transformations (\ref{1.12})  with themselves:
    \begin{equation}
    \begin{aligned}
    &(\delta_{\lambda_1}\delta_{\lambda_2}-\delta_{\lambda_2}\delta_{\lambda_1})q^{+}_a=\frac{1}{8}(\lambda^{i}_{2a}\lambda^{b}_{1i}-\lambda^{i}_{1a}\lambda^b_{2i})q^{+}_{b}-\frac{1}{8}(\lambda^{ia}_2\lambda_{1aj}-\lambda^{ia}_1\lambda_{2aj})\left(u^{+}_i\frac{\partial}{\partial u^{+}_j}+u^{-}_i\frac{\partial}{\partial u^{-}_j}\right)q^{+}_{a},\\
    &(\delta_{\lambda_1}\delta_{\lambda_2}-\delta_{\lambda_2}\delta_{\lambda_1})W=-\frac{1}{8}(\lambda^{ia}_2\lambda_{1ja}-\lambda^{ia}_1\lambda_{2ja})\left(u^{+}_i\frac{\partial}{\partial u^{+}_j}+u^{-}_i\frac{\partial}{\partial u^{-}_j}\right)W.
    \end{aligned}
    \end{equation}
    The details of the derivation are given in Appendix. We observe that the on-shell commutator of two hidden $R$-symmetry transformations yields
    manifest linear $SU(2)_R$ and $SU(2)_{\rm PG}$ transformations, as should be.

\subsection{Effective action}
    The part of the superfield $\mathcal{ N} = 1$ SYM effective action containing the component four-derivative
    term of the gauge field reads \cite{litlink 11}
    \begin{equation}
    S_0=c_0\int d^{13}zdu\, W \ln \frac{W}{\Lambda},
    \label{1.13}
    \end{equation}
    where $W$ is the abelian gauge superfield strength, $\Lambda$ is a scale parameter and $c_0$ is a
    dimensionless constant.

    The variation of    action (\ref{1.13}) under the transformation (\ref{1.12}) is as follows
    \begin{equation}
    \begin{split}
    &\delta S_0=c_0\int d^{13}zdu \ln W \delta W\\
    &=\frac{c_0}{4}\int d^{13}zdu \ln W \left(2\lambda^{+a}q^{-}_{a}-2\lambda^{-a}q^{+}_{a}
    +\lambda^{-a}\theta^{+\hat\alpha}D^{+}_{\hat \alpha} q^{-}_a-\lambda^{+a}\theta^{-\hat \alpha}D^{+}_{\hat \alpha} q^{-}_a\right)\\
    &=-\frac{c_0}{4}\int d^{13}zdu \frac{\lambda^{-a}\theta^{+\hat\alpha}D^{+}_{\hat \alpha}W q^{-}_a-\lambda^{+a}\theta^{-\hat \alpha}D^{+}_{\hat \alpha}W q^{-}_a}{W}.
    \label{1.14}
    \end{split}
    \end{equation}
    It can be partially canceled by variation of the extra term
    \begin{equation}
    S_1=c_1\int d^{13}zdu \frac{q^{+a}q^{-}_a}{W}.
    \label{1.15}
    \end{equation}
    The variation of (\ref{1.15}) reads
    \begin{equation}
    \begin{split}
        &\delta S_1=c_1\delta\int d^{13}zdu \frac{q^{+a}q^{-}_a}{W}\\
        &=-c_1\int d^{13}zdu \frac{\left(\lambda^{+a}  W-\lambda^{+a}\theta^{-\hat\alpha}D^{+}_{\hat \alpha}  W
            +\lambda^{-a}\theta^{+\hat \alpha}D^{+}_{\hat \alpha} W \right)q^{-}_a}{W}\\
        &-\frac{c_1}{4}\int d^{13}zdu \frac{q^{+b}q^{-}_b}{W^2}\left(2\lambda^{+a}q^{-}_{a}
        -2\lambda^{-a}q^{+}_{a}+\lambda^{-a}\theta^{+\hat\alpha}D^{+}_{\hat \alpha} q^{-}_a-\lambda^{+a}\theta^{-\hat \alpha}D^{+}_{\hat \alpha} q^{-}_a\right).
    \end{split}
    \label{1.16}
    \end{equation}
    Due to the property $\int d^{13}zdu\ \lambda^{+a} q^{-}_a=0$, the first term in (\ref{1.16}) exactly cancels (\ref{1.14}),
    provided that $c_1=-c_0/4$.

    Hence,
    \begin{equation}
        \begin{split}
        &\delta(S_0+S_1)=-\frac{c_1}{4}\int d^{13}zdu \frac{q^{+b}q^{-}_b}{W^2}\left[2\lambda^{+a}q^{-}_{a}-2\lambda^{-a}q^{+}_{a}+\left(\lambda^{-a}\theta^{+\hat\alpha}
        -\lambda^{+a}\theta^{-\hat \alpha}\right)D^{+}_{\hat \alpha} q^{-}_a\right]\\
        &=-\frac{c_1}{4}\int d^{13}zdu \frac{q^{+b}q^{-}_b}{W^2}\left[4\lambda^{+a}q^{-}_{a}+(\lambda^{-a}\theta^{+\hat\alpha}
        -\lambda^{+a}\theta^{-\hat \alpha})D^{+}_{\hat \alpha} q^{-}_a\right].
        \end{split}
        \label{1.17}
    \end{equation}
    Consider the last term here in more detail:
    \begin{equation}
    \begin{split}
    &-\frac{c_1}{4}\int d^{13}zdu \frac{q^{+b}q^{-}_b}{W^2}(\lambda^{-a}\theta^{+\hat\alpha}-\lambda^{+a}\theta^{-\hat \alpha})D^{+}_{\hat \alpha} q^{-}_a\\
    & =-\frac{c_1}{4}\int d^{13}zdu \bigg[\frac{-4\lambda^{+a}W +2(\lambda^{-a}\theta^{+\hat\alpha}-\lambda^{+a}\theta^{-\hat \alpha})D^{+}_{\hat \alpha} W}{W^3}
    q^-_a(q^{+b}q^{-}_b)\\
    &-(\lambda^{-a}\theta^{+\hat\alpha}-\lambda^{+a}\theta^{-\hat \alpha})\frac{q^{+b}D^{+}_{\hat \alpha} q^{-}_bq^{-}_a}{W^2}\bigg]
    \\
    &
    =-\frac{c_1}{4}\int d^{13}zdu \bigg[\frac{-4\lambda^{+a}W +2
        (\lambda^{-a}\theta^{+\hat\alpha}-\lambda^{+a}\theta^{-\hat \alpha})D^{+}_{\hat \alpha} W}{W^3}q^-_a(q^{+b}q^{-}_b)\\&
        -(\lambda^{-a}\theta^{+\hat\alpha}-\lambda^{+a}\theta^{-\hat \alpha})\frac{q^{+b} q^{-}_bD^{+}_{\hat \alpha}q^{-}_a}{2W^2}\bigg].
    \end{split}
    \end{equation}
    We used  the integration by parts with respect to the spinor derivative in the second line and
    cyclic identities for $SU(2)$ indices  in the third line. We observe that the last term in the third line equals the
    expression we started with, but with he minus sign. Hence,
   \begin{equation}
   \begin{split}
   &-\frac{c_1}{4}\int d^{13}zdu \frac{q^{+b}q^{-}_b}{W^2}(\lambda^{-a}\theta^{+\hat\alpha}-\lambda^{+a}\theta^{-\hat \alpha})D^{+}_{\hat \alpha} q^{-}_a\\
    &= -\frac{c_1}{3}\int d^{13}zdu \frac{[-2\lambda^{+a}W +
       (\lambda^{-a}\theta^{+\hat\alpha}-\lambda^{+a}\theta^{-\hat \alpha})D^{+}_{\hat \alpha} W]q^-_a(q^{+b}q^{-}_b)}{W^3}.
    \end{split}
    \label{3.22}
    \end{equation}
    Substituting (\ref{3.22}) into (\ref{1.17}), we obtain
        \begin{equation}
    \begin{split}
    \delta(S_0+S_1)&=-\frac{c_1}{3}\int d^{13}zdu\frac{\left(\lambda^{+}_a  W-\lambda^{+}_a\theta^{-\hat\alpha}D^{+}_{\hat \alpha}  W
        +\lambda^{-}_a\theta^{+\hat \alpha}D^{+}_{\hat \alpha} W \right)q^{-}_a(q^{+b}q^{-}_b)}{W^3}.
    \end{split}
    \label{1.18}
    \end{equation}

    Once again,  the variation of (\ref{1.18}) can be partially canceled by the variation of the additional term
    \begin{equation}
    S_2=c_2\int d^{13}zdu \,\frac{(q^{+a}q^{-}_a)^2}{W^3}, \qquad c_2=-\frac{c_1}{6}.
    \end{equation}

    Finally, we consider the general expression
    \begin{equation}
    S=\int d^{13}zdu\left[ W \ln \frac{W}{\Lambda} +\sum\limits^{\infty}_{n=1}c_n\frac{(q^{+a}q^{-}_a)^n}{W^{2n-1}}\right].
    \label{1.19}
    \end{equation}
    One can show that this expression is invariant under  the transformation (\ref{1.12}), if
    \begin{equation}
        -(n+1)c_{n+1}=\frac{n(2n-1)}{n+2}c_n.
    \end{equation}

    Therefore, the action (\ref{1.19}) is equal to
    \begin{equation}
    S=c_0\int d^{13}zdu W \left[\ln \frac{W}{\Lambda} +\frac{1}{2}H(Z)\right],
    \label{3.28}
    \end{equation}
    where
    \begin{equation}
        Z=\frac{q^{+a}q^{-}_a}{W^2},
    \end{equation}
    and
    \begin{equation}
    H(Z)=1+2\ln\frac{1+\sqrt{1+2Z}}{2}+\frac{2}{3}\frac{1}{1+\sqrt{1+2Z}}-\frac{4}{3}\sqrt{1+2Z}.
    \label{3.30}
    \end{equation}
    This expression coincides with the one obtained in \cite{litlink 5} by resorting to
    hidden supersymmetry instead of $R$-symmetry\footnote{Recently,
    the expression (\ref{3.30}) has been derived by the direct quantum calculation \cite{BIM-1}.}.

    One can also directly verify that (\ref{3.28}) in invariant under the $R$-symmetry transformations
    \begin{equation}
    \begin{split}
    &\delta S=c_0\int d^{13}zdu\Big[\frac{(\lambda^{+a}\theta^{-\hat \alpha}-\lambda^{-a}\theta^{+\hat\alpha})D^{+}_{\hat \alpha}W q^{-}_a}{4W}\\
    &-\frac{H'(Z)\left[\lambda^{+a}  W-(\lambda^{+a}\theta^{-\hat\alpha}
        -\lambda^{-a}\theta^{+\hat \alpha})D^{+}_{\hat \alpha} W \right]q^{-}_a}{2W}\\
    &+\frac{1}{8}[H(Z)-2ZH'(Z)]\left[4\lambda^{+a}q^{-}_{a}+(\lambda^{-a}\theta^{+\hat\alpha}-\lambda^{+a}\theta^{-\hat \alpha})D^{+}_{\hat \alpha} q^{-}_a\right]\Big].
    \end{split}
    \end{equation}
    This expression can be simplified  with the help of the identity
    \begin{equation}
    \begin{split}
    &\int dz^{13}du\, [H(Z)-2ZH'(Z)]\left[4\lambda^{+a}q^{-}_{a}+(\lambda^{-a}\theta^{+\hat\alpha}-\lambda^{+a}\theta^{-\hat \alpha})D^{+}_{\hat \alpha} q^{-}_a\right] \\
    & = \int dz^{13}du\ 2[1+2H'(Z)] \bigg[(\lambda^{-a}\theta^{+\hat\alpha}-\lambda^{+a}\theta^{-\hat \alpha}) \frac{q^{-}_aD^{+}_{\hat \alpha}W}{W}+\lambda^{+a}q^{-}_a\bigg],
    \end{split}
    \end{equation}
    which is derived by employing the integration by parts with respect to the spinor derivative and using the
    definition of the function $H(Z)$ (\ref{3.30}).

    Thus we obtain
    \begin{equation}
    \begin{split}
    &\delta S=c_0\int d^{13}zdu\bigg[\frac{(\lambda^{+a}\theta^{-\hat \alpha}-\lambda^{-a}\theta^{+\hat\alpha})D^{+}_{\hat \alpha}W q^{-}_a}{4W}\\
    &-\frac{H'(Z)\left[\lambda^{+a}  W-(\lambda^{+a}\theta^{-\hat\alpha}
        -\lambda^{-a}\theta^{+\hat \alpha})D^{+}_{\hat \alpha} W \right]q^{-}_a}{2W}\bigg]\\
    &+c_0\int dz^{13}du\, \frac{2}{8}\,  [1+2H'(Z)] \bigg[(\lambda^{-a}\theta^{+\hat\alpha}-\lambda^{+a}\theta^{-\hat \alpha})
    \frac{q^{-}_aD^{+}_{\hat \alpha}W}{W}+\lambda^{+a}q^{-}_a\bigg]=0\,.
    \end{split}
    \end{equation}

We conclude that the condition of invariance under $R$-symmetry can be employed instead
of the demand of hidden supersymmetry in order to construct the complete $5D, \,\mathcal{ N}=2$ invariant
superspace functional, starting from the functional which is invariant  under the manifest $\mathcal{ N}=1$
supersymmetry only.

\section{Summary}
In this paper we have found the realization of $R$-symmetry for
$4D, \,\mathcal{ N}=4$ and  $5D, \,\mathcal{ N}=2$ supersymmetric
gauge theories as the superfield transformations in the relevant
harmonic superspaces. The $R$-symmetry transformations were defined in
the explicit form, and they mix the gauge multiplet and hypermultiplet
harmonic superfields with each other. It was proved that the microscopic actions
of $4D$, $\mathcal{ N}=4$ and  $5D$, $\mathcal{ N}=2$ SYM theories
are invariant under these transformations without any on-shell conditions on the superfields involved. Thus, the above
transformations constitute an additional invariance of  $4D$,
$\mathcal{ N}=4$ and  $5D$, $\mathcal{ N}=2$ supersymmetric gauge
theories.

The algebraic structure of the harmonic superfield $R$-symmetry
transformations was studied. First, these transformations form the
closed algebra only on shell. Second, the $R$-symmetry transformations
are consistent with both manifest and hidden supersymmetry transformations,
which are a necessary element of the harmonic superspace formulations of the maximally extended
SYM theories. To be more precise, the $R$-symmetry transformations
form a closed algebra with the manifest and hidden supersymmetry
transformations. This means, in particular, that the $R$-symmetry
transformations and the hidden supersymmetry transformations are in a sense interchangeable. If a manifestly invariant superfield
functional is invariant under the $R$-symmetry transformations, then it
will be automatically invariant under the hidden supersymmetry
transformations and vise versa.

The $R$-symmetry transformations were applied to the problem of
the hypermultiplet completion of  the low-energy effective action of $4D$,
$\mathcal{ N}=4$ and $5D$, $\mathcal{ N}=2$ SYM theories, proceeding from the low-energy effective actions
in the gauge multiplet sector. Using
these transformations, we constructed the leading low-energy complete
effective actions for the theories just mentioned, beginning with the terms
containing only the gauge multiplet contributions. We have shown that
the hypermultiplet dependence of the effective actions under
consideration is completely specified by the requirement of
invariance under the $R$-symmetry transformations. We focused on the
case of $SU(2)$ gauge group spontaneously broken to $U(1)$.
A generalization to other gauge groups is straightforward. An interesting
property  is that the effective action is not only invariant under
the $R$-symmetry transformations but can be fixed by them up to an overall constant.

It would be tempting to reveal other possible implications of hidden $R$-symmetry
in extended superfield gauge theories in diverse dimensions. The maximally supersymmetric gauge theory
in six-dimension is ${\cal N}=(1,1)$ SYM theory. It has only manifest linear
$SU(2)_R\times SU(2)_{\rm PG}$ $R$-symmetry and for this reason the methods of the present paper
seem not to be appropriate for analysis of the structure of the quantum effective action of this theory in the ${\cal N}= (1,0)$
harmonic superspace formulation. Only the considerations based on the hidden ${\cal N}=(0,1)$ supersymmetry prove to be adequate \cite{BIM1,BIM2}.
On the other hand, the hidden $R$-symmetry method could be useful in the harmonic superspace formulations of
$6D, {\cal N}= (2,0)$ tensor multiplet (see, e.g., \cite{BuPl}). Indeed, only $SU(2)_R$ $R$-symmetry is manifest there, while the rest
of the full $USp(4)$ $R$-symmetry of $6D,\, {\cal N}= (2,0)$ supersymmetry should be realized as a hidden symmetry. We plan to consider
this and some other additional examples of exploiting superfield hidden $R$-symmetries elsewhere.

\section*{Acknowledgments}
\noindent  I.L.B. and E.A.I.
are grateful to RFBR grant, project No. 18-02-01046, for a partial
support.

\appendix
\section{Appendix}
    \subsection{Commutators of $R$-symmetry transformations in $4D, \mathcal{ N}=4$ SYM theory}
        In this section we directly calculate the commutator of $R$-symmetry with itself for $4D$, $\mathcal{ N}=4$ SYM theory (\ref{3333}).

    Let us start with the transformation of $\bar W$
    \begin{equation}
    \begin{split}
    &(\delta_{\lambda_1}\delta_{\lambda_2}-\delta_{\lambda_2}\delta_{\lambda_1})\bar W
    =\frac{1}{2}\delta_{\lambda_1}\left[\lambda^{-a}_2q^{+}_{a}-\lambda^{+a}_2q^{-}_{a}
    -\left(\lambda^{-a}_2\theta^{+\alpha}-\lambda^{+a}_2\theta^{- \alpha}\right)D^{+}_{ \alpha} q^{-}_a\right]\\
    &-(\lambda_1 \leftrightarrow \lambda_2).\\
    \end{split}
    \end{equation}
    The right-hand side of this relation can further be worked out as
    \begin{equation}
    \begin{split}
   & \frac{1}{8}\bigg[\lambda^{-a}_2\left[ \lambda^{+}_{1a}  W
    +\left(
    \lambda^{-}_{1a}\theta^{+\alpha}-\lambda^{+}_{1a}\theta^{-\alpha}\right)D^{+}_\alpha W+\bar\lambda^{+}_{1a}{\bar W}
    + \left(\bar \lambda^{-}_{1a}\bar\theta^{+\dot\alpha}
    -\bar \lambda^{+}_{1a}\bar\theta^{-\dot\alpha}\right)\bar D^{+}_{\dot\alpha}{\bar W}
    \right]
    \\
    &-\lambda^{+a}_2
    \left[ \lambda^{-}_{1a}  W+
    \left(\lambda^{-}_{1a}\theta^{+\alpha}-\lambda^{+}_{1a}\theta^{-\alpha}\right)D^{-}_\alpha W+\bar\lambda^{-}_{1a}{\bar W}
    + \left(\bar \lambda^{-}_{1a}\bar\theta^{+\dot\alpha}
    -\bar \lambda^{+}_{1a}\bar\theta^{-\dot\alpha}\right)\bar D^{-}_{\dot\alpha}{\bar W}
    \right]\\
    &
    +(\lambda^{-a}_2\theta^{+\alpha}-\lambda^{+a}_2\theta^{- \alpha})
    \big[
    \lambda^{-}_{1a}D^{+}_{ \alpha}  W-\lambda^{+}_{1a}D^{-}_{ \alpha}  W-
    \left(\lambda^{-}_{1a}\theta^{+\beta}-\lambda^{+}_{1a}\theta^{-\beta}\right)D^{+}_{ \alpha} D^{-}_\beta W\\
    &- \left(\bar \lambda^{-}_{1a}\bar\theta^{+\dot\alpha}
    -\bar \lambda^{+}_{1a}\bar\theta^{-\dot\alpha}\right)D^{+}_{ \alpha} \bar D^{-}_{\dot\alpha}{\bar W}
    \big]
    \bigg]
    -(\lambda_1 \leftrightarrow \lambda_2).
    \end{split}
    \label{44.1}
    \end{equation}
    To properly transform this expression, we note first that the full coefficient before
    $D^+_\alpha D^{-}_\beta W $ in the next-to-last line in (\ref{44.1})
    is proportional to $\epsilon^{\alpha\beta}$,
    while $D^{+ \alpha}D^{-}_{\alpha}W= 0$, as follows from the equation of motion (\ref{2.44}) for $W$ and the constraints (\ref{01.12}).
    Analogously, using the relations (\ref{01.11}), we can replace $D^{+}_{ \alpha} \bar D^{-}_{\dot\alpha}{\bar W}$ in the last line with $-2i\partial_{\alpha\dot \alpha}\bar W$.
    In addition, we introduce the notations
    \begin{equation}
    \lambda^{ij}=\lambda^{(ia}_2\bar \lambda^{j)}_{1a}-\lambda^{(ia}_1\bar \lambda^{j)}_{2a}, \qquad \lambda=\lambda^{ia}_2\bar \lambda_{1ia}-\lambda^{ia}_1\bar \lambda_{2ia},
    \end{equation}
    where $\lambda^{ij}$ refer to $SU(2)_R$ transformations and $\lambda$ ($\bar \lambda=-\lambda$)
    to $U(1)_R$ transformations. As a result we obtain
\begin{equation}
    \begin{split}
    &(\delta_{\lambda_1}\delta_{\lambda_2}-\delta_{\lambda_2}\delta_{\lambda_1})\bar W\\
    &
    =\frac{1}{8}\bigg[-\lambda \bar W+ \frac{1}{2}\lambda \bar \theta^{-\dot \alpha}\bar D^{+}_{\dot \alpha}\bar W
    - \frac{1}{2}\lambda \bar \theta^{+\dot \alpha}\bar D^{-}_{\dot \alpha}\bar W- \frac{1}{2}\lambda  \theta^{+\dot \alpha} D^{-}_{ \alpha}\bar W\\
    &-2i\lambda\left(\theta^{+\alpha}\bar \theta^{-\dot \alpha}
    -\theta^{-\alpha}\bar \theta^{+\dot \alpha}\right)\partial_{\alpha{\dot \alpha}}\bar W\\
    &+(\lambda^{--}\bar \theta^{+\dot \alpha}-\lambda^{-+}\bar \theta^{-\dot \alpha})\bar D^{+}_{\dot \alpha}\bar W
    -(\lambda^{+-}\bar \theta^{+\dot \alpha}-\lambda^{++}\bar \theta^{-\dot \alpha})\bar D^{-}_{\dot \alpha}\bar W
    \\
    &-(\lambda^{+-} \theta^{+ \alpha}-\lambda^{++} \theta^{- \alpha}) D^{-}_{ \alpha}\bar W\\
    &-2i\left(\lambda^{--}\theta^{+\alpha}\bar \theta^{+\dot \alpha}-\lambda^{-+}\theta^{+\alpha}\bar \theta^{-\dot \alpha}
    -\lambda^{+-}\theta^{-\alpha}\bar \theta^{+\dot \alpha}
    +\lambda^{++}\theta^{-\alpha}\bar \theta^{-\dot \alpha}\right)\partial_{\alpha{\dot \alpha}}\bar W
    \bigg]
    \\
      &
    =\frac{1}{8}\bigg[ \frac{1}{2}\lambda \bar \theta^{-\dot \alpha}\frac{\partial}{\partial\bar\theta^{-\dot{\alpha}}}\bar W
    + \frac{1}{2}\lambda \bar \theta^{+\dot \alpha}\frac{\partial}{\partial\bar\theta^{+\dot{\alpha}}}\bar W
    - \frac{1}{2}\lambda  \theta^{+ \alpha}\frac{\partial}{\partial\theta^{+{\alpha}}}\bar W\\
    &+(\lambda^{--}\bar \theta^{+\dot \alpha}-\lambda^{-+}\bar \theta^{-\dot \alpha})\frac{\partial}{\partial\bar\theta^{-\dot{\alpha}}}\bar  W
    +(\lambda^{+-}\bar \theta^{+\dot \alpha}-\lambda^{++}\bar \theta^{-\dot \alpha})\frac{\partial}{\partial\bar\theta^{+\dot{\alpha}}}\bar W
    \\
    &+(\lambda^{+-} \theta^{+ \alpha}-\lambda^{++} \theta^{- \alpha})\frac{\partial}{\partial\theta^{+{\alpha}}}\bar W\\
    &-2i\left(\lambda^{--}\theta^{+\alpha}\bar \theta^{+\dot \alpha}-\lambda^{-+}\theta^{+\alpha}\bar \theta^{-\dot \alpha}
    +\lambda^{+-}\theta^{+\alpha}\bar \theta^{-\dot \alpha}
    -\lambda^{++}\theta^{-\alpha}\bar \theta^{-\dot \alpha}\right)\partial_{\alpha{\dot \alpha}}\bar W
    \bigg].
    \end{split}
    \label{44.2}
    \end{equation}
Due to the relations (\ref{01.12}), this expression can be cast in the form
\begin{equation}
    \begin{split}
    (\delta_{\lambda_1}\delta_{\lambda_2}-\delta_{\lambda_2}\delta_{\lambda_1})\bar W
    =&-\frac{1}{8}\lambda^{i}_j\left(u^{+}_i\frac{\partial}{\partial u^{+j}}+u^{-}_i\frac{\partial}{\partial u^{-j}}\right)\bar W\\
    &+\frac{1}{8}\bigg[\frac{1}{2}\lambda \bar \theta^{-\dot \alpha}\frac{\partial}{\partial\bar\theta^{-\dot{\alpha}}}\bar W
    + \frac{1}{2}\lambda \bar \theta^{+\dot \alpha}\frac{\partial}{\partial\bar\theta^{+\dot{\alpha}}}\bar W
    + \frac{1}{2}\bar \lambda  \theta^{+ \alpha}\frac{\partial}{\partial\theta^{+{\alpha}}}\bar W-\lambda \bar W
    \bigg].
    \end{split}
    \label{44.3}
    \end{equation}
    The transformation law of $W$ can  be obtained through the complex conjugation.

     Next, we pass to the transformation of $q^{+}_a$
     \begin{equation}
    \begin{split}
    &(\delta_{\lambda_1}\delta_{\lambda_2}-\delta_{\lambda_2}\delta_{\lambda_1})q^{+}_a=\frac{1}{2}\delta_{\lambda_1}\bigg[ \lambda^{+}_{2a}  W
    +\left(
    \lambda^{-}_{2a}\theta^{+\alpha}-\lambda^{+}_{2a}\theta^{-\alpha}\right)D^{+}_\alpha W\\
    &+\bar\lambda^{+}_{2a}{\bar W}
    + \left(\bar \lambda^{-}_{2a}\bar\theta^{+\dot\alpha}
    -\bar \lambda^{+}_{2a}\bar\theta^{-\dot\alpha}\right)\bar D^{+}_{\dot\alpha}{\bar W}
    \bigg]-(\lambda_1\leftrightarrow\lambda_2)
    \\
    &
    =\frac{1}{8}\bigg[\lambda^{+}_{2a}
    \left[\bar\lambda^{-b}_1q^{+}_{b}-\bar\lambda^{+b}_1q^{-}_{b}-\left(\bar\lambda^{-b}_1 \bar\theta^{+\dot\alpha}
    -\bar\lambda^{+b}_1\bar \theta^{-\dot \alpha}\right)\bar D^{+}_{\dot \alpha} q^{-}_b\right]
    \\
    &+\bar \lambda^{+}_{2a}   \left[\lambda^{-b}_1q^{+}_{b}-\lambda^{+b}_1q^{-}_{b}-\left(\lambda^{-b}_1\theta^{+\alpha}
    -\lambda^{+b}_1\theta^{- \alpha}\right)D^{+}_{ \alpha} q^{-}_b\right]\\
    &+  \left(\lambda^{-}_{2a}\theta^{+\alpha}-
    \lambda^{+}_{2a}\theta^{- \alpha}\right)
    \left[-\bar\lambda^{+b}_1D^{+}_{ \alpha}q^{-}_{b}+\left(\bar\lambda^{-b}_1 \bar\theta^{+\dot\alpha}
    -\bar\lambda^{+b}_1\bar \theta^{-\dot \alpha}\right)D^{+}_{ \alpha}\bar D^{+}_{\dot \alpha} q^{-}_b\right]\\
    &+\left(\bar\lambda^{-}_{2a}\bar\theta^{+\dot\alpha}-
    \bar\lambda^{+}_{2a}\bar\theta^{- \dot\alpha}\right)
    \left[-\lambda^{+b}_1\bar D^{+}_{ \dot\alpha}q^{-}_{b}+\left(\lambda^{-b}_1 \theta^{+\alpha}
    -\lambda^{+b}_1 \theta^{- \alpha}\right)\bar D^{+}_{\dot \alpha} D^{+}_{ \alpha} q^{-}_b\right]
    \bigg]-(\lambda_1 \leftrightarrow \lambda_2).
    \label{44.6}
    \end{split}
    \end{equation}
    Thanks to the analyticity of $q^{+}_a$ and the equation of motion (\ref{2.44}), one can replace
    $\bar D^{+}_{\dot \alpha} D^{+}_{ \alpha} q^{-}_b$ in the last line of (\ref{44.6}) with $-2i\partial_{\alpha\dot \alpha}q^{+}_b$, and similarly
    for $ D^{+}_{ \alpha}\bar  D^{+}_{ \dot \alpha} q^{-}_b$. The Lie bracket parameter
    \begin{equation}
    \lambda^{ab}_{\text{(PG)}}=\lambda^{i(a}_{2}\bar \lambda^{b)}_{1i}-\lambda^{i(a}_{1}\bar \lambda^{b)}_{2i}
    \end{equation}
    is just associated with the $SU(2)_{\text{PG}}$ symmetry transformations.
Therefore,
    \begin{equation}
    \begin{split}
    &
    (\delta_{\lambda_1}\delta_{\lambda_2}-\delta_{\lambda_2}\delta_{\lambda_1})q^{+}_a \\
    &
    =\frac{1}{8}\bigg[\lambda^b_{\text{(PG)}a}q^{+}_b-\lambda^{+-}q^{+}_a+\lambda^{++}q^{-}_a
    +\frac{1}{2}\lambda\bar \theta^{+\dot\alpha}\frac{\partial}{\partial \bar \theta^{+\dot \alpha}}q^{+}_a
    -\frac{1}{2}\lambda \theta^{+\alpha}\frac{\partial}{\partial  \theta^{+ \alpha}}q^{+}_a\\
    &+\left(\lambda^{+-}\bar \theta^{+\dot\alpha}-\lambda^{++}\bar \theta^{-\dot \alpha}\right)
    \frac{\partial}{\partial \bar \theta^{+\dot \alpha}}q^{+}_a+\left(\lambda^{-+} \theta^{+\alpha}
    -\lambda^{++} \theta^{- \alpha}\right)\frac{\partial}{\partial  \theta^{+ \alpha}}q^{+}_a
    \\
    &+2i\left(-\lambda^{--}\theta^{+\alpha}\bar\theta^{+\dot\alpha}+\lambda^{+-}\theta^{-\alpha}\bar\theta^{+\dot\alpha}
    +\lambda^{-+}\theta^{+\alpha}\bar\theta^{-\dot\alpha}-\lambda^{++}\theta^{-\alpha}\bar\theta^{-\dot\alpha}
    \right)\partial_{\alpha{\dot \alpha}}q^{+}_a
    \bigg].
    \label{44.8}
    \end{split}
    \end{equation}
        Using the equation of motions (\ref{2.44}), this expression can be brought to the form
        \begin{equation}
        \begin{split}
            (\delta_{\lambda_1}\delta_{\lambda_2}-\delta_{\lambda_2}\delta_{\lambda_1})q^{+}_a&
            =\frac{1}{8}\left[\lambda^{b}_{\text{(RG)}a}q^{+}_b+\frac{1}{2}\lambda\bar \theta^{+\dot\alpha}\frac{\partial}{\partial \bar \theta^{+\dot \alpha}}q^{+}_a
            +\frac{1}{2}\bar\lambda \theta^{+\alpha}\frac{\partial}{\partial  \theta^{+ \alpha}}q^{+}_a\right]\\
            &-\frac{1}{8}\lambda^{i}_j\left(u^{+}_i\frac{\partial}{\partial u^{+j}}+u^{-}_i\frac{\partial}{\partial u^{-j}}\right)q^{+}_a,
        \end{split}
        \end{equation}
        where $\lambda$ is the bracket parameter for $U(1)_{{R}}$ transformations.

    \subsection{Commutators of $R$-symmetry transformations in $5D, \mathcal{ N}=2$ SYM theory}
    In this section we calculate the commutator of $R$-symmetry with itself for $5D$, $\mathcal{ N}=2$
    SYM theory (\ref{1.12}).

    We start with the transformation of $W$
        \begin{equation}
        \begin{split}
        &
        (\delta_{\lambda_1}\delta_{\lambda_2}-\delta_{\lambda_2}\delta_{\lambda_1})W
        =\frac{1}{4}\delta_{\lambda_1}\Big[2(\lambda^{+a}_2q^{-}_{a}-\lambda^{-a}_2q^{+}_{a})
        +\left(\lambda^{-a}_2\theta^{+\hat\alpha}-\lambda^{+a}_2\theta^{-\hat \alpha}\right)D^{+}_{\hat \alpha} q^{-}_a\Big]\\
        &  -(\lambda_1 \leftrightarrow \lambda_2).
        \end{split}
        \end{equation}
        Its right-hand side is evaluated to be
        \begin{equation}
        \begin{split}
        &-\frac{1}{8}\bigg[
        -(\lambda^{+a}_2\lambda^{+}_{1a}-\lambda^{+a}_1\lambda^{+}_{2a})\theta^{-\hat\alpha}D^{-}_{\hat \alpha}W+(\lambda^{+a}_2\lambda^{-}_{1a}
        -\lambda^{+a}_1\lambda^{-}_{2a})\theta^{+\hat\alpha}D^{-}_{\hat \alpha} W\\ &+(\lambda^{-a}_2\lambda^{+}_{1a}
        -\lambda^{-a}_1\lambda^{+}_{2a})\theta^{-\hat\alpha}D^{+}_{\hat \alpha} W-(\lambda^{-a}_2\lambda^{-}_{1a}
        -\lambda^{-a}_1\lambda^{-}_{2a})\theta^{+\hat\alpha}D^{+}_{\hat \alpha} W
        \\
        &
        +\big[(\lambda^{-a}_2\theta^{+\hat\alpha}-\lambda^{+a}_2\theta^{-\hat \alpha})(\lambda^{+}_{1a}\theta^{-\hat \beta}
        -\lambda^{-}_{1a}\theta^{+\hat \beta})\\
        &-(\lambda^{-a}_1\theta^{+\hat\alpha}-\lambda^{+a}_1\theta^{-\hat \alpha})(\lambda^{+}_{2a}\theta^{-\hat \beta}
        -\lambda^{-}_{2a}\theta^{+\hat \beta})\big]D^{+}_{\hat \alpha}D^{-}_{\hat \beta} W \bigg].
        \end{split}
        \label{5.1}
        \end{equation}
        To bring this expression to a simpler form, we make use of  the relation
        \begin{equation}
        \big(D^{+}_{\hat \alpha}D^{-}_{\hat \beta}+
        D^{-}_{\hat \alpha}D^{+}_{\hat \beta}\big) W=0
        \end{equation}
        which follows from the equation of motion (\ref{3.9}) for $W$, the constraint (\ref{3.10}) and the definition (\ref{3.2}) of $W$\footnote{One also needs
        to use the identity $\Omega_{\hat{\alpha}\hat{\beta}} = \frac12 \varepsilon_{\hat{\alpha}\hat{\beta}\hat{\gamma}\hat{\nu}}\Omega^{\hat{\gamma}\hat{\nu}}$.}.
         Using this relation, we can
        replace $D^{+}_{\hat \alpha}D^{-}_{\hat \beta} $ with $i\partial_{\hat \alpha\hat \beta}$ in the last line of (\ref{5.1})
        due to  the antisymmetry of the full factor in front of it with respect to the indices $\hat \alpha$ and $\hat \beta$. We also introduced the notation
        \begin{equation}
        \lambda^{ij}=\lambda^{ia}_2\lambda^{j}_{1a}-\lambda^{ia}_1\lambda^{j}_{2a}\,.
        \end{equation}

        Rewriting eq. (\ref{5.1}) and substituting the explicit expressions for $D^{-}_{\hat \alpha} $ and $D^{+}_{\hat \alpha} $, we obtain
        \begin{equation}
        \begin{split}
            &(\delta_{\lambda_1}\delta_{\lambda_2}-\delta_{\lambda_2}\delta_{\lambda_1})W\\
            &
        =-\frac{1}{8}\bigg[
        \left(\lambda^{++}\theta^{-\hat\alpha}-\lambda^{+-}\theta^{+\hat\alpha}\right)\frac{\partial W}{\partial \theta^{+\hat \alpha}}
        +2i\left(\lambda^{++}\theta^{-\hat\alpha}-\lambda^{+-}\theta^{+\hat\alpha}\right)\theta^{-\hat \gamma}\partial_{\hat \alpha\hat \gamma}W\\
         &+\left(\lambda^{-+}\theta^{-\hat\alpha}-\lambda^{--}\theta^{+\hat\alpha}\right)\frac{\partial W}{\partial \theta^{-\hat \alpha}}
        \\
        &+\left(i\lambda^{-+}\theta^{+\hat \alpha}\theta^{-\hat \beta}+i\lambda^{+-}\theta^{-\hat \alpha}\theta^{+\hat \beta}
        -i\lambda^{--}\theta^{+\hat \alpha}\theta^{+\hat \beta}
        -i\lambda^{++}\theta^{-\hat \alpha}\theta^{-\hat \beta}\right)\partial_{\hat \alpha\hat \beta}W\bigg]
        \\
        &
        =-\frac{1}{8}\bigg[
        \left(\lambda^{++}\theta^{-\hat\alpha}-\lambda^{+-}\theta^{+\hat\alpha}\right)\frac{\partial W}{\partial \theta^{+\hat \alpha}}
        +\left(\lambda^{-+}\theta^{-\hat\alpha}-\lambda^{--}\theta^{+\hat\alpha}\right)\frac{\partial W}{\partial \theta^{-\hat \alpha}}
        \\
        &+i(\lambda^{-+}\theta^{-\hat \alpha}
        -\lambda^{--}\theta^{+\hat \alpha})\theta^{+\hat \beta}\partial_{\hat \alpha\hat \beta}W
        +i(\lambda^{++}\theta^{-\hat \alpha}-\lambda^{+-}\theta^{+\hat \alpha})\theta^{-\hat \beta}\partial_{\hat \alpha\hat \beta}W\bigg]
        \\
        &=
        -\frac{1}{8}\lambda^{i}_{j}\left(u^{+}_i\frac{\partial}{\partial u^{+}_j}+u^{-}_i\frac{\partial}{\partial u^{-}_j}\right)W.
        \end{split}
    \end{equation}
    When passing  to the last line, we exploited the relations  (\ref{3.10}).

    Now we proceed to the transformations of $q^{+}_a$,
    \begin{equation}
    \begin{split}
    &(\delta_{\lambda_1}\delta_{\lambda_2}-\delta_{\lambda_2}\delta_{\lambda_1})q^{+}_a
    =-\frac{1}{2}\delta_{\lambda_1}\bigg[\lambda^{+}_{2a}  W-\lambda^{+}_{2a}\theta^{-\hat\alpha}D^{+}_{\hat \alpha}  W
    +\lambda^{-}_a\theta^{+\hat \alpha}D^{+}_{\hat \alpha} W \bigg]-(\lambda_1 \leftrightarrow \lambda_2)\\
    &
    =-\frac{1}{8}\bigg[\lambda^{+}_{2a}  \left(2\lambda^{+b}_1q^{-}_{b}-2\lambda^{-b}_1q^{+}_{b}
    +\lambda^{-b}_1\theta^{+\hat\alpha}D^{+}_{\hat \alpha} q^{-}_b-\lambda^{+b}_1\theta^{-\hat \alpha}D^{+}_{\hat \alpha} q^{-}_b\right)\\
    &-\left(\lambda^{+}_{2a}\theta^{-\hat\alpha}-
    \lambda^{-}_{2a}\theta^{+\hat \alpha}\right)D^{+}_{\hat \alpha}\left(2\lambda^{+b}_1q^{-}_{b}+(\lambda^{-b}_1\theta^{+\hat\beta}
    -\lambda^{+b}_1\theta^{-\hat \beta})D^{+}_{\hat \beta} q^{-}_b\right)\bigg]-(\lambda_1 \leftrightarrow \lambda_2)\\
    &
    =-\frac{1}{8}\bigg[
    -\left(\lambda^{i}_{2a}\lambda^{b}_{1i}-\lambda^{i}_{1a}\lambda^{b}_{2i}\right)q^{+}_b-\left(\lambda^{+b}_{2}\lambda^{+}_{1b}
    -\lambda^{+b}_{1}\lambda^{+}_{2b}\right)q^{-}_{a}
    +\left(\lambda^{+b}_{2}\lambda^{-}_{1b}-\lambda^{+b}_{1}\lambda^{-}_{2b}\right)q^{+}_{a}
    \\
    &-\lambda^{+b}_2\left(\lambda^{+}_{1b}\theta^{-\hat\alpha}-
    \lambda^{-}_{1b}\theta^{+\hat \alpha}\right)D^{-}_{\hat \alpha}q^{+}_a
    +\lambda^{+b}_1\left(\lambda^{+}_{2b}\theta^{-\hat\alpha}-
    \lambda^{-}_{2b}\theta^{+\hat \alpha}\right)D^{-}_{\hat \alpha}q^{+}_a\\
    &-2i\left(\lambda^{+b}_{2}\theta^{-\hat\alpha}+
    \lambda^{-b}_{2}\theta^{+\hat \alpha}\right)\left(\lambda^{+}_{1b}\theta^{-\hat\beta}-
    \lambda^{-}_{1b}\theta^{+\hat \beta}\right)\partial_{\hat\alpha\hat\beta}q^{+}_a\bigg],
    \end{split}
    \end{equation}
    where the relations $D^{-}_{\hat{\alpha}}q^{+a}=-D^{+}_{\hat{\alpha}}q^{-a}$ and
    $D^{+}_{\hat \alpha}D^{+}_{\hat{\beta}}q^{-a}=-2i\partial_{\hat \alpha\hat \beta}q^{+a}$ were used.

    Substituting the explicit expressions for $D^{-}_{\hat \alpha}$, we finally obtain
    \begin{equation}
    \begin{split}
    &(\delta_{\lambda_1}\delta_{\lambda_2}-\delta_{\lambda_2}\delta_{\lambda_1})q^{+}_a \\
    &
    =-\frac{1}{8}\bigg[
    -\left(\lambda^{i}_{2a}\lambda^{b}_{1i}-\lambda^{i}_{1a}\lambda^{b}_{2i}\right)q^{+}_b-
    \left(\lambda^{+b}_{2}\lambda^{+}_{1b}-\lambda^{+b}_{1}\lambda^{+}_{2b}\right)q^{-}_{a}
    +\left(\lambda^{+b}_{2}\lambda^{-}_{1b}-\lambda^{+b}_{1}\lambda^{-}_{2b}\right)q^{+}_{a}
    \\
    &+\lambda^{+b}_2\left(\lambda^{+}_{1b}\theta^{-\hat\alpha}-
    \lambda^{-}_{1b}\theta^{+\hat \alpha}\right)\frac{\partial q^{+}_a}{\partial \theta^{+\hat \alpha}}
    -\lambda^{+b}_1\left(\lambda^{+}_{2b}\theta^{-\hat\alpha}-
    \lambda^{-}_{2b}\theta^{+\hat \alpha}\right)\frac{\partial q^{+}_a}{\partial \theta^{+\hat \alpha}}\\
    &-2i\lambda^{+b}_1\lambda^{+}_{2b}\theta^{-\hat\alpha}\theta^{-\hat\beta}\partial_{\hat \alpha\hat \beta}q^{+}_a
    -2i\lambda^{-b}_1\lambda^{-}_{2b}\theta^{+\hat\alpha}\theta^{+\hat\beta}\partial_{\hat \alpha\hat \beta}q^{+}_a\bigg]
    \\
    &=-\frac{1}{8}\left(\lambda^{ia}_2\lambda_{1aj}-\lambda^{ia}_1\lambda_{2aj}\right)\left(u^{+}_i
    \frac{\partial}{\partial u^{+}_j}+u^{-}_i\frac{\partial}{\partial u^{-}_j}\right)q^{+}_{a}+\frac{1}{8}\left(\lambda^{i}_{2a}\lambda^{b}_{1i}-\lambda^{i}_{1a}\lambda^b_{2i}\right)q^{+}_{b}.
    \end{split}
    \end{equation}
When passing to the last line, we used the equation of motion  (\ref{3.9}).

\end{document}